\documentclass[12pt]{article}

\newcounter{multieqs}

\newcommand{\bq}{\begin{equation}}
\newcommand{\fq}{\end{equation}}
\newcommand{\bqr}{\begin{eqnarray}}
\newcommand{\fqr}{\end{eqnarray}}
\newcommand{\non}{\nonumber \\}

\newcommand{\com}[2]{[ #1 , #2 ]}

\newcommand{\rf}[1]{(\ref{#1})}

\def\alp{\alpha}   \def\bet{\beta}    \def\gam{\gamma}
\def\del{\delta}   \def\eps{\epsilon} 
         
       \def\lam{\lambda} 
 \def\sig{\sigma}

\def\Gam{\Gamma}   \def\Del{\Delta}   \def\The{\Theta}

\def\cD{{\cal D}}  \def\cF{{\cal F}}

\def\cM{{\cal M}}  \def\cO{{\cal O}}

\def\pa{\partial} 

\def\be{\begin{equation}}
\def\ee{\end{equation}}
\def\bea{\begin{eqnarray}}
\def\eea{\end{eqnarray}}

\def\pr{^{\prime}}

\def\rar{\rightarrow}

\def\one{1\!\!1\,\,}

\newcommand{\tr}{\mbox{Tr}}

\def\hlf{\frac{1}{2}}
\def\ove#1{\frac{1}{#1}}

\def\Tr{\,{\mbox{Tr}}}
\def\STr{\,{\mbox{STr}}}
\def\Sym{\,{\mbox{Sym}}}
\def\Xd{\dot{X}}
\def\Zd{\dot{Z}}
\def\Xdd{\ddot{X}}
\def\dt{\frac{d}{dt}}

\begin{document}

\def\titel{
\thispagestyle{empty}
\setcounter{page}{0}


\begin{flushright}
\begin{tabular}{l}
ITFA-2001-23 \\
NIKHEF-H/01-010 \\
hep-th/0108161  \\ 
\end{tabular}
\end{flushright}

\vspace{9mm}
\begin{center}

{\Large \bf 
General covariance of the non-abelian DBI-action}\\
\vspace{12mm}

{Jan de Boer\footnote{E-mail:
jdeboer@science.uva.nl}~~and~
Koenraad Schalm\footnote{E-mail: kschalm@nikhef.nl}
}

\vskip 1cm

{${}^{1}$ {\em Institute for Theoretical Physics\\ 
University of Amsterdam \\
Valckenierstraat 65\\
1018 XE Amsterdam, The Netherlands}} \\[5mm]
{${}^{2}$ {\em NIKHEF Theory Group\\
P.O. Box 41882 \\
1009 DB Amsterdam, The Netherlands}} \\[5mm]

\vspace{5mm}

{\bf Abstract}
\end{center}
In this paper we study the action for $N$ D0-branes in a curved
background. In particular, we focus on the meaning of space-time
diffeomorphism invariance. For a single D-brane, diffeomorphism
invariance acts in a naive way on the world-volume fields, 
but for multiple D-branes, the meaning of diffeomorphism invariance
is much more obscure. The problem goes beyond the determination of an
ordering of the $U(N)$-valued fields, because
one can show that there is no lift of ordinary diffeomorphisms
to matrix-valued diffeomorphisms. On the other hand, the action
can presumably be constructed from perturbative string theory 
calculations. Based on the general characteristics of such
calculations we determine a set of constraints on the 
action for $N$ D0-branes, that ensure space-time covariance. 
These constraints can be solved order by order, but
they are
insufficient to determine the action completely. 
All solutions to the constraints obey the axioms of D-geometry. 
Moreover the action must contain new terms. This exhibits clearly that
the answer is more than a suitable 
ordering of the action of a single D0 brane.

\vfill
\newpage
\setcounter{footnote}{0}
}
\titel

\section{Introduction}  

In string theory there is no invariant notion of 
space-time geometry, rather it depends in a non-trivial way
on the object that is being used to probe the geometry.
In particular, closed strings see a different
geometry from the one seen by open strings or the
one seen by D-branes. As explained in \cite{douglas},
the metric seen by a single D0-brane is a well-defined quantity 
and is referred to by the term `D-geometry'. It is an
interesting question what the geometry is that is seen
by $N$ coincident D-branes. 
The flat-space action for $N$ coincident
 D-branes is given by
the dimensional reduction of $N=1$ $d=10$ $U(N)$ Super-Yang-Mills, so
already in this case something non-trivial happens. The
adjoint scalar fields now act as probes of the surrounding
geometry; as 
space-time coordinates have been replaced by $N\times N$ matrices, it
suggests that space-time has become non-commutative
\cite{Witten:1996im}.
 
The generalization of this action to curved space is
the problem we wish to address in this paper. This is a
special case of the more general problem to find
the action for D-branes in arbitrary closed string
backgrounds. In the presence of e.g. a closed string
NS-NS B-field, the action for D-branes remains similar except
that the ordinary product of fields is replaced by
a non-commutative $\ast$-product \cite{douglashull}.
Thus, a non-zero $B$-field gives rise to new 
geometrical structures on the D-brane world volume.
The presence of transversal RR fields gives rise to
several interesting physical phenomena, such as the Myers
effect \cite{myers}, which resolves classical supergravity
singularities. One may expect that similar phenomena will
appear in the description of D-branes in curved space, i.e.
that on the one hand some new non-commutative generalization
of curved space will appear, and that on the other hand
we will find a purely geometrical version of the Myers 
effect.

The approach we will take is to study the notion of diffeomorphism
invariance for D0-branes. If we know how the world-volume fields
transform under space-time diffeomorphisms, the action in a 
curved background is completely determined, as 
the graviton is the gauge field
enforcing local coordinate invariance.
In particular,
the kinetic term in the action of a single D0 brane, $S \sim
\int \! dt \, g_{\mu\nu} \dot{x}^{\mu} \dot{x}^{\nu}$, is completely
fixed by diffeomorphism invariance. If we replace the fields
$x$ by $N\times N$ matrices, we can write down a similar
action where the metric is replaced by a matrix-valued metric,
and the group of diffeomorphisms is replaced by a group
of matrix-valued diffeomorphisms. The matrix-valued metric
should be a functional of the original metric, since both
actions are derived in the same closed string background.
In particular, since the closed string background is
diffeomorphism invariant, the same group of diffeomorphisms
should be represented both in the action of a single as
well as in the action for multiple D-branes. On the other hand, we will
show that there is no lift of the diffeomorphism group into
the group of matrix-valued diffeomorphisms. It follows that a new
symmetry is needed to impose diffeomorphism invariance for
matrix-valued fields. We will propose such a symmetry and
explore its 
consequences. Though requirement of diffeomorphism invariance
considerably constrains the action for matrix-valued fields, it does
not uniquely determine it. This non-uniqueness is consistent
with the results of \cite{Okawa:2001sh,Okawa:2001if}, where
different results for the stress-tensors of D-branes in 
bosonic and superstring theory were obtained.

The action for D-branes in curved space has been considered before
by various authors. In \cite{douglas,kato} the
idea was to find an ordering of the answer for a single D-brane
that has the property that the masses of off-diagonal fluctuations
of the matrix-valued fields have masses proportional to the
geodesic length between separated D-branes. In \cite{dec,apr} 
the linear coupling to the background metric was derived using
the matrix theory interpretation \cite{banks} of the action of
$N$ D0-branes. The linear coupling to the background metric
can also be derived directly using worldsheet methods and
is related to the stress-tensor of non-commutative gauge theories
\cite{Okawa:2001if,Liu:2001ps,Das:2001ur}. Alternatively,
one can try to use the connection with matrix theory to
try to determine the coupling \cite{plefka}. 
Through T-duality the
D0-brane action is related to the Non-abelian Born-Infeld action. A promising
approach here is the matching of its BPS spectrum with that of 
string theory \cite{Brecher:1998tv}. Symmetry principles
have also been used e.g. in \cite{Bergshoeff:2000ik} in an
attempt to find the ordering of the terms that are higher order
in the gauge field field strength. For some specific curved
backgrounds, one is able to explicitly construct the action, see 
\cite{Berenstein:1998dw} amongst others.

The outline of this paper is as follows.
In section 2 we give a precise formulation of the problem.
We show that the linear results of \cite{dec,apr} are invariant
under space-time diffeomorphisms up to terms of higher order.
In section 3 we show that 
the group of matrix-valued diffeomorphisms does not
contain regular diffeomorphisms in a straightforward way. 
Using input from open-string scattering
amplitudes we reformulate the full non-linear problem into
a set of consistency
conditions on the manifestly covariant normal coordinate version of
the theory. In section 4 we give the solution to these constraints
through sixth order in the fields and show the necessity of
novel vertices. 
In section~5 we discuss some properties of the action.
We propose a curved-space generalization of the 
${\rm Tr}([X^i,X^j]^2)$ potential in terms of the matrix-valued
metric that appears in the kinetic term. With this potential,
which is covariant and reproduces the linearized coupling
of \cite{dec,apr}, the action
automatically fulfills the
axioms of D-geometry put forth by Douglas.
Conclusions are given in section~6, and some details that
are omitted from the main text are given in the appendices.

We will adhere to the convention that all upper-case
``coordinates'' $X,Y,Z,\ldots$ are matrix-valued and do not commute,
whereas 
lower-case ``coordinates'' $x,y,z,\ldots$ are regular commuting
ones. Roman indices $i,k,\ell, \ldots$ indicate directions transverse
to the D-brane, Greek indices $\mu,\nu,\ldots$ target-space
directions (worldvolume plus transverse) and the Greek indices from the
beginning of the alphabet $\alp,\bet,\gam,\ldots$ are $U(N)$ matrix-labels.

\section{The Problem}

\subsection{Covariance and Particle-Geometry}

As we briefly mentioned in the introduction, the effective action 
for the zero-modes of a single D$p$-brane in flat
space, up to second order in derivatives, 
is given by the reduction of $N=1$ $d=10$ abelian
super-Yang-Mills to $p+1$ dimensions. The scalars obtained from the
dimensional reduction parametrize the transverse motion. 
For a D0-brane the action ---  a gauged linear sigma model --- 
is therefore simply that of a point particle in nine Euclidean 
dimensions upon elimination of the non-dynamical $U(1)$ gauge field. 
\begin{equation}
  \label{eq:1}
  S_{D0}^{free} = -\frac{T_0}{2}\int d\tau D_{\tau}x^{i}D_{\tau}x_{i} 
~~~~\stackrel{
A_{\tau}=0}
{\longrightarrow} 
~~~S_{D0}^{free} = -\frac{T_0}{2}\int d\tau \dot{x}^{i}\dot{x}_{i} 
\end{equation}
with $D_{\tau}=\pa/\pa\tau+A_{\tau}$.

Requiring in addition that the action is  
invariant under coordinate changes $\del
x^{i} = \xi^{i}(x)$ in the
nine transverse directions, yields, of course, the action for a
massive particle 
in a curved background with mass $T_0$.
\begin{equation}
  \label{eq:2}
  S_{D0} = -\frac{T_0}{2}\int d\tau
  g_{ij}(x)\dot{x}^{i}\dot{x}^{j}  
\end{equation}
Let us briefly recall the details why.

With a local infinitesimal diffeomorphism
\begin{equation}
\del x^i = \xi^i(x)
\end{equation}
one derives the gravitational current, the stress tensor,
\begin{equation}
T^{ij}(x) = \frac{T_0}{2}\int
d\tau \del^D(x-x(\tau))\dot{x}^i\dot{x}^j~. 
\label{str}
\end{equation}
It is conserved on shell
\begin{equation}
\pa_iT^{ij}(x) = \frac{T_0}{2}\int d\tau \del^D(x-x(\tau))\ddot{x}^j = -
\int d\tau \del^D(x-x(\tau))\frac{\del S_{D0}^{free}}{\del x_j} 
\end{equation}
for the flat-space particle action. The stress
tensor thus
couples naturally to the linearized gauge-field, the graviton
$h_{ij}=g_{ij}-\eta_{ij}$. 
\begin{equation}
S_{D0} = S_{D0}^{free} + \int d^Dx T^{ij}(x) h_{ij}(x)~.
\label{1part}
\end{equation}
The action  (\ref{1part}) is to first order invariant under linearized gauge
transformations (diffeomorphisms) where the graviton transforms into the
derivative of the local gauge parameter. 
\begin{eqnarray}
\del x^i &=& \xi^i(x) \non 
\del h_{ij} &=& -\pa_{(i}\xi_{j)} \non
\del S_{D0} &=& 0 + \cO(\xi^2,\xi h,h^2) 
\end{eqnarray}
To find an action invariant to all orders, one uses e.g. the Noether
method. One finds one only needs to adjust the transformation rules of
$h_{ij}$ to include non-linear terms, to the standard transformations
of a metric
\begin{equation}
  \label{eq:4}
  \del(\eta_{ij}+h_{ij}) \equiv \del g_{ij}= -D_{(i}\xi_{j)}
\end{equation}
with $D_{i}$ the covariant derivative with the Christoffel connection
of $g_{ij}$. There is no need to change
the transformation rules of $x$ nor to add additional couplings to the
action. 

As the integrations in \rf{1part} are over two different sets of
parameters it is more convenient to think of the stress tensor in
terms of moments 
\begin{eqnarray}
\int d^Dx T^{ij}(x) h_{ij}(x) &=&  \sum_{n=0}^{\infty} \int
\frac{d^Dx}{n!} T^{ij}(x)x^{k_1}\ldots x^{k_n}\pa_{k_1}\ldots \pa_{k_n}
h_{ij}(0) \non 
&=&  \sum_{n=0}^{\infty} \int \frac{d\tau}{n!} T^{ij(k_1\ldots
k_n)}(x(\tau))\pa_{k_1}\ldots \pa_{k_n} h_{ij}(0)~. 
\end{eqnarray}
In the last step we have used the delta function in
\rf{str}. This clarifies that in quantum field theory terms the
particle action is better understood as belonging to an infinite set
of equivalent theories related by the (coupled global) ``symmetry''
transformations on the fields $x^i$ and couplings $g_{ij}$  
\begin{eqnarray}
Z(g_{ij}) = \int \cD x \exp (S_{part}[x^i,g_{ij}]) & \simeq & \int
\cD (x+\del x) \exp( S_{part}[ x^i+\del x^i,g_{ij}+\del g_{ij}])\non  
\del x^i &=& \sum_{n=0}^{\infty} \ove{n!}x^{k_1}\ldots
x^{k_n}\pa_{k_1}\ldots \pa_{k_n} \xi^i(0)\non 
\del \pa_{k_1}\ldots \pa_{k_n}h_{ij}(0) &=& -
\sum_{n=0}^{\infty}\pa_{k_1}\ldots \pa_{k_n}\pa_{(i} \xi_{j)}(0) 
\end{eqnarray}
Except for the ``symmetries'' with $n=0,1$ all of these are
non-linear. 

\subsection{Formulation of the problem: \\
Covariance and D-geometry}

For $N$ D0-branes, superposed in flat space, the gauge group enhances
to $U(N)$ and becomes
non-abelian \cite{Witten:1996im}. The 
scalars $x^{i}(\tau)$ become $N \times N$ matrices transforming in 
the adjoint. The
immediate connection with the location of the D-brane in space-time is
therefore lost. Only to the (diagonal) expectation values can we ascribe
such a meaning. The question we wish to answer is how to
couple this theory to a curved background.

The obvious generalization of (\ref{eq:2}) is
\begin{equation}
  \label{eq:3}
  S_{D0} = -\frac{T_0}{2}\int d\tau\, \tr\, G_{i\alp\bet\, j\gam\del}(X)
  \dot{X}^{i\alp\bet}\dot{X}^{j\gam\del} .
\end{equation}
which is the most general action one can write down to second
order in the derivatives. We included explicitly the matrix indices
$\alp,\bet,\gam,\del$ in the action. It is clear that this action is 
invariant under general ``matrix-valued'' diffeomorphisms, {\em if} we
view the triple $i\alp\bet $ as a single index and
we let the metric transform in the usual way. Eq. (\ref{eq:3}) is 
nothing but an unconventional form of the sigma model action for $dN^2$
degrees of freedom ($i=1\ldots d; \alp,\bet=1\ldots N$) and the starting
point of D-geometry \cite{douglas}. In section 4.4 we will take this
action as a starting point to construct a diffeomorphism invariant 
effective action for $N$
D-branes in $d$ dimensions.

What are the further requirements the action (\ref{eq:3}) must
fulfill (cf. the axioms of D-geometry \cite{douglas})? 
First of all, 
the action can be derived from string theory disc diagrams
with arbitrarily many graviton vertex operators in the
interior and scalar field vertex operators
carrying Chan-Paton factors on the boundary. Hence the action should
consist of a single-trace \cite{Tseytlin:1997cs}. 
In other words, if we would expand $G$
around $X^i=0$ in a power series, the action should be 
a sum of traces of polynomials in $X^{i}$. 

A second requirement, which is also obvious from string theory,
is that if take the matrices $X^{i}$ to be diagonal, the
action should reduce to a sum of (\ref{eq:1}) for each
diagonal entry, i.e. a sum of $N$ actions for a single D0 brane.

The third requirement of \cite{douglas} is that if we 
  expand the action around a diagonal configuration for 
$X^{i}$, the
masses of the off-diagonal fluctuations should be proportional
to the geodesic distance between the points given by the
diagonal values of $X^{i}$. This assumes that the off-diagonal
elements of $X^i$ correspond to the vertex operators that 
create open strings between separated D-branes. 
These masses are independent of field redefinitions that
preserve the single trace structure of the action, and will
therefore be diffeomorphism invariant. To determine the
masses we need the 
curved space version of the potential term ${\rm Tr}([X,X]^2)$
that is part of the full non-abelian D0-brane action. 
We postpone the construction of the potential and the 
calculation of the masses to section~5.

The fourth axiom of D-geometry:
  that the classical moduli space is the 
  symmetric product $\cM^N/S_N$ of $N$
  copies of the transverse space $\cM$ also depends on the exact form
  of the potential and will be discussed in this section as well.

\bigskip

The main important new ingredient is that we will explicitly
  impose diffeomorphism invariance, in addition
to the four axioms of \cite{douglas}. To state this requirement,
we need to remember that the metric in (\ref{eq:3}) is not
arbitrary but a functional of the closed string metric $g_{ij}$.
In other words, the action (\ref{eq:3}) is really an functional
of $g_{ij}$ and $X^i$,
\be S \equiv S[g_{ij},X^{j}]. \ee
Now suppose that $g$ and $g'$ are two metrics related by a target
space diffeomorphism. Physics should not be sensitive to the
difference between $g$ and $g'$, and therefore there should
exist a field redefinition $X \rightarrow X'(x)$ such that
\be \label{eq:3b}
S[g,X] = S[g',X'] .
\ee
As we will show, this is in fact 
a highly non-trivial requirement.

{Finally we will also impose a sixth requirement} that the 
linearized coupling of the
graviton in the action agrees with the coupling proposed in
\cite{dec,apr}, and which has been confirmed by world-sheet
calculations in \cite{Okawa:2001if}. The result of \cite{dec,apr}
is that at the linearized level the coupling is the
completely symmetrized trace. We briefly review this coupling
below and verify that it is consistent with space-time
diffeomorphism invariance. Imposing this constraint means 
that
our results are compatible with the linear coupling of the
metric to D-branes in superstring theory. If we would be
interested in the coupling of D-branes in bosonic string theory
to a curved background, we should  
change this sixth requirement to the
appropriate form of the linearized coupling found there
\cite{Okawa:2001if}.

\subsubsection{The linearized coupling}

In \cite{dec,apr} Taylor and van Raamsdonk put forward a proposal for 
the coupling of the
linearized graviton to the action for $N$ free D0-branes. Moreover, 
the stress tensor proposed is conserved on
shell \cite{vr}. This, as for the single D0-brane, must mean that the
variation of the action under an 
infinitesimal local transformation can be cancelled by requiring that
the gauge field coupling to the conserved current varies into the
derivative of the parameter. Therefore, the D0-brane action of
\cite{dec} with 
the first order coupling of the stress-tensor to the linearized
graviton,  
\begin{equation}
S = -\frac{T_0}{2} \int d\tau \tr\left(\dot{X}^i\dot{X}_i\right) +
\int d\tau \sum_{n=0}^{\infty}T^{ij(k_1\ldots k_n)} {\pa_{k_1}\ldots
\pa_{k_n}}h_{ij}(0)  
\label{actlin}
\end{equation}
where
\begin{equation}
T^{ij(k_1\ldots k_n)} =
\frac{T_0}{2n!}\mbox{Str}\left(\dot{X}^i\dot{X}^jX^{k_1}\ldots X^{k_n}\right) ~,
\end{equation}
must in fact be invariant under linearized infinitesimal
diffeomorphisms. As has become standard, we use here the symbol {\bf
  Str} 
for the symmetrized 
trace: the trace over the completely symmetrized ordering of
the matrices.

Indeed, it is not so hard to check, assuming the moments of
$h_{ij}$ transform in the standard way, that the action is invariant
under such linearized diffeomorphisms 
\begin{eqnarray}
\del h_{ij}(x) &=& \pa_{(i}\xi_{j)}(x) ~~\rar~~    
\del \pa_{k_1}\ldots \pa_{k_n}h_{ij}(0)  = \pa_{k_1}\ldots
\pa_{k_n}\pa_{(i}\xi_{j)}(0) \\ 
\del X^i &=& \sum_{n=0}^{\infty} \ove{n!}X^{k_1}\ldots
X^{k_n}\pa_{k_1}\ldots \pa_{k_n}\xi^{i}(0) = \xi^{i}(X)~.
\end{eqnarray}  
Thus, at the linearized level space-time diffeomorphisms are
realized as certain
``matrix-valued''
diffeomorphisms. Note that there is no ``unexpected'' ordering in the
transformation 
for $X^i$. A priori, this is not the only possible answer. The 
$n$-th derivative of the parameter $\xi$ can only be contracted with
the totally symmetric combination, but one could have internal
contractions in addition. 
To agree with the $N \rar 1$ limit such terms must
involve commutators.  

In the above we have ignored the potential term proportional to
$[X,X]^2$ 
as well
as the fermions required by supersymmetry. It is just as easy to show,
that the action including these terms, is also invariant under the above
transformations. Of course, when we include fermions we should
change the metric coupling to the vielbein. For
completeness, the action 
\begin{eqnarray} 
S&=& \frac{T_0}{2} \int d\tau \tr \left( \dot{X}^i\dot{X}^i +
\hlf\com{X^i}{X^j} 
\com{X^i}{X^j} + i \The \dot{\The} - \The\Gam_i\com{X^i}{\The}\right)
\non 
&&+ \sum_{n=0}^{\infty}\int d\tau T^{i(k_1\ldots k_n)}_a\pa_{k_1}\ldots
\pa_{k_n}e_{i}^a ~,\\ 
T^{i(k_1\ldots k_n)}_a &=&-
\frac{T_0}{2n!}\mbox{Str}\left[\left(2\dot{X}^i\dot{X}^je^{(0)}_{ja} -
2\com{X^i}{X^p} 
\com{X^p}{X^j}e^{(0)}_{ja} -
\The\Gam_a\com{X^i}{\The}\right)X^{k_1}\ldots X^{k_n}\right]~, 
\end{eqnarray}
where $e^{(0)}_{ja}=\del_{ja}$ is the zeroth order vielbein, is
invariant under 
\begin{eqnarray}  
\del X^i &=& 
 \sum_{n=0}^{\infty} \ove{n!}X^{k_1}\ldots X^{k_n}\pa_{k_1}\ldots \pa_{k_n}\xi^{i}(0) = \xi^{i}(X) ~,\non
\del e_{i}^a(x) &=& -\pa_{i}\xi_{a}(x) ~~~~~~~~~\Leftrightarrow\non    
\del \pa_{k_1}\ldots \pa_{k_n}e_{i}^a(0)  &=& -\pa_{k_1}\ldots \pa_{k_n}\pa_{i}\xi_{a}(0) ~.\label{potferm}
\end{eqnarray}
As we said above, we will require that the linearized coupling
of the metric is as in (\ref{actlin}).

For a large part we will suppress the potential and the fermions in
the remainder of the paper and return to them in section 5.3.

Note that there is a different way to write (\ref{actlin}), namely
\be
S=-\frac{T_0}{2} \int dt \int dk \int_0^1 d\lambda g_{mn}(k) 
{\rm Tr}( e^{i \lambda k_{j} X^{j} } \dot{X}^{m}
 e^{i (1-\lambda) k_{\ell} X^{\ell} } \dot{X}^{n} )
\ee
where $g_{mn}(k)$ is the Fourier transform of the metric.
This is the form that is obtained from world-sheet calculations
\cite{Okawa:2001if}, and is reminiscent of the open Wilson lines
that appear in the couplings of closed string fields to
non-commutative gauge theories.

\medskip

In summary, 
starting with a sigma model of $dN^2$ degrees of freedom, we
want to find a $d$-dimensional (gauged) nonlinear sigma model built from
 $N \times N$ 
matrix-valued fields that is invariant under target space
 diffeomorphisms ({\em a}). This action must in addition
\begin{itemize}
\item[({\em b})] consist of a single group trace,
\item[({\em c})] reduce to $N$-copies of the standard particle
  action if we take the matrices $X$ to be diagonal.
\item[({\em d})] yield masses proportional to the geodesic distance for
  off-diagonal fluctuations,
\item[({\em e})] and reduce to the action (\ref{actlin}) 
found by Taylor and van Raamsdonk, at the linearized level in the metric. 
\end{itemize}
Requirements ({\em b})-({\em d}) are part of the axioms of D-geometry;
({\em a}) is new, and 
({\em e}) is ''experimental input''.

\section{Structure of Matrix-valued diffeomorphisms}

As is expressed in (\ref{eq:3b}), given two metrics $g,g'$
that are related by a space-time diffeomorphism, there
should exist a field redefinition of the $N \times N$ matrix-coordinates 
$X'(X)$ such that the
action remains invariant. These field redefinitions implement
space-time diffeomorphism invariance, in terms of certain
specific matrix-valued coordinate changes.

Unlike regular diffeomorphisms the specific redefinitions $X'(X)$
that implement space-time diffeomorphism invariance are not
completely arbitrary, but it is not a priori clear what
$X'(X)$ should be. It definitely should be compatible with
the single trace structure, and therefore $X'$ should
be a function of $X$ without explicit traces, e.g. a power
series. 
Naively, one would expect that $X'(X)$ would be a suitable
ordered form of $x'(x)$. We now show that there cannot be
such a unique ordering, and as a consequence that the group of
matrix-valued diffeomorphisms does not contain the group
of space-time diffeomorphisms as a subgroup.

To explain this, let us assume for the moment that we have solved the
ordering problem and consider a change of
coordinates $x \rightarrow x' \rightarrow x''$, with corresponding
metrics $g,g',g''$. Then there must be field redefinitions 
$X'(X)$, $X''(X')$ such that 
\begin{equation}
 \label{e1}
S[g,X] = S[g',X'(X)] = S[g'',X''(X'(X))].
\end{equation}
On the other hand, by the group property, 
we can also consider $x \rightarrow x''$ directly,
leading to $S[g,X] = S[g'',X''(X)]$. Combining with the previous equation
(\ref{e1}) we find that 
\begin{equation} \label{e2}
S[g'',X''(X'(X))] = S[g'',X''(X)].
\end{equation}
Now the only obvious invariance of the action $S$ is the $U(N)$ gauge
invariance, so it hard to believe that (\ref{e2}) can be true unless
\begin{equation} \label{e3}
X''(X'(X)) = U X''(X) U^{\dagger}
\end{equation}
with $U \in U(N)$. (In other words, the metric $G_{i \alp\bet,j \gam \del}$ has
in general a $U(N)$ group worth of Killing vectors, but no more).
This means the following. Let us denote by
{\bf diff} the regular group of diffeomorphisms and by {\bf
  DIFF}
the group of diffeomorphisms of the $N \times N$ matrix-valued
coordinates $X$, compatible with the single trace structure. Eq. 
(\ref{e3}) shows that in order for the above scenario to work 
there should be a projective representation of {\bf diff} in {\bf
  DIFF}, projective with respect to $U(N)$. 

It is, however, straightforward to show that such a representation does
not exist.  
Consider two infinitesimal transformations
\begin{eqnarray}
x'^{i}  & = & x^{i} + a^{i}_{pq} x^{p} x^{q} +
     b^{i}_{pqr} x^{p} x^{q} x^{r}
  + {\cal O}(x^4) \\{}
x''^{i}  & = & x'^{i} + c^{i}_{pq} x'^{p} x'^{q} +
     d^{i}_{pqr} x'^{p} x'^{q} x'^{r}
  + {\cal O}(x'^4) \\ \nonumber
  & = & x^{i} + (a^{i}_{pq}+c^{i}_{pq}) x^{p}
 x^{q} +(b^{i}_{pqr} + 2 c^{i}_{ps}
 a^{s}_{qr} + d^{i}_{pqr}) x^{p} 
 x^{q} x^{r}
  + {\cal O}(x'^4)
\end{eqnarray}
and corresponding matrix redefinitions
\begin{eqnarray} \label{jjj1}
X'^{i}  & = & X^{i} + F^{i}_{pq}(a) X^{p} X^{q} +
     G^{i}_{pqr}(a,b) X^{p} X^{q} X^{r}
  + {\cal O}(X^4) \\{}
X''^{i}  & = & X'^{i} + F^{i}_{pq}(c) X'^{p} X'^{q} +
     G^{i}_{pqr}(c,d) X'^{p} X'^{q} X'^{r}
  + {\cal O}(X'^4) \label{jjj2}
\end{eqnarray}
where $a,b,c,d$ are symmetric in the lower indices but $F,G$ are not
necessarily symmetric. In order to work order by order, the
function  
$F$ can be a function of $a (\mbox{or}~c)$ only, but $G$ can be a
function of both $a,b$ (or $c,d$), etc. The functions $F$ and $G$ tell
us how to order 
the terms
in order to lift a diffeomorphism from {\bf diff} to {\bf DIFF}. 
By inserting (\ref{jjj2}) into (\ref{jjj1}) we obtain
\begin{eqnarray}
  \label{eq:7}
  X''^{i}(X'(X)) &=& 
  X^{i}+\left(F^{i}_{pq}(a)+F^{i}_{pq}(c)\right)
  X^{p}X^{q}+\non
&+&
  \left(G^{i}_{pqr}(a,b)+G^{i}_{pqr}(c,d) + F^{i}_{ps}(c)F^{s}_{qr}(a)+F^{i}_{sr}(c)F^{s}_{pq}(a)\right)X^{p}X^{q}X^{r}.
\end{eqnarray}
This should equal (schematically)
\be \label{eq:7a}
X''^{i}(X)=U( X^{i} + F^{i}_{pq}(a+c) X^{p} X^{q} +
     G^{i}_{pqr}(a+c,b+d+2 c a) X^{p} X^{q} X^{r}
  + {\cal O}(X^4))U^{\dagger} .
\ee
To lowest order we see that $U=1$. 
A generic ansatz for $F$ and $G$ is
\bea
F^{i}_{pq}(a) & = & a^{i}_{pq} \\
G^{i}_{pqr}(a,b) & = & 
b^{i}_{pqr} + \lambda_1 a^{i}_{ps}
a^{s}_{qr} + \lambda_2 a^{i}_{qs}
a^{s}_{rp} + (-\lambda_1 -\lambda_2)
a^{i}_{rs} a^{s}_{pq} .
\eea
If we insert this in (\ref{eq:7}) and (\ref{eq:7a}) and demand
that the equations are identical, we obtain 
\bea
& &  c^{i}_{ps} a^{s}_{qr} +
c^{i}_{sr} a^{s}_{pq} 
+  \lambda_1 a^{i}_{ps}
a^{s}_{qr} + \lambda_2 a^{i}_{qs}
a^{s}_{rp} + (-\lambda_1 -\lambda_2)
a^{i}_{rs} a^{s}_{pq} \nonumber \\
&&+  \lambda_1 c^{i}_{ps}
c^{s}_{qr} + \lambda_2 c^{i}_{qs}
c^{s}_{rp} + (-\lambda_1 -\lambda_2)
c^{i}_{rs} c^{s}_{pq} \nonumber \\
&=&  \lambda_1 (a+c)^{i}_{ps}
(a+c)^{s}_{qr} + \lambda_2 (a+c)^{i}_{qs}
(a+c)^{s}_{rp} + (-\lambda_1 -\lambda_2)
(a+c)^{i}_{rs} (a+c)^{s}_{pq} \nonumber \\
& & +\frac{2}{3} (  c^{i}_{ps} a^{s}_{qr} +
c^{i}_{sr} a^{s}_{pq} +
c^{i}_{sq} a^{s}_{rp}) .
\eea
We immediately see that there is no solution for $\lambda_1$, $\lambda_2$
such that this equation holds. It is possible to make a more general
argument that does not rely on a specific form of $F$ and $G$, but
we omit that here for simplicity.

The fact that {\bf diff} is not a subgroup\footnote{Notice that 
{\bf diff} is however a quotient of {\bf DIFF}, namely by
the group of matrix-valued diffeomorphisms that are of commutator type, 
which are the diffeomorphisms that
reduce to the identity when restricted to diagonal matrices. }
 of {\bf DIFF} shows
that in the equation (\ref{eq:3b}),
\be 
\label{eq:3bagain}
S[g,X]=S[g'(g),X'(X)], \ee
the relation between the matrix-valued diffeomorphism $X'(X)$ and
the ordinary diffeomorphism that relates $g$ and $g'$ is
not straightforward. 

A different way to reach the same 
conclusion is to start with the linearized form
of $S[g,X]$ given in (\ref{actlin}). If we require a symmetry
of the form (\ref{eq:3bagain}) we can use the Noether
procedure to deduce the higher order couplings to the 
metric. Given an action consisting of a field $\phi$ and 
a gauge
field $A$, the 
assumption that it is invariant under some specific variations $\del
\phi$ and $\del A$  requires the variation should vanish 
order by order in $\eps$, where $\eps$  counts
powers of the variation parameter and the gauge field,
\begin{eqnarray}
\del \phi &=& \eps \del^{(1)}\phi+\eps^2\del^{(2)}\phi +  \ldots \non
\del A  &=& \del^{(0)}A+\eps\del^{(1)}A + \eps^2\del^{(2)}A+ \ldots \non
S &=& S^{(0)} + \eps^1S^{(1)} + \eps^2S^{(2)}+\ldots  ;~~~~~~~~\rar  \frac{\pa}{\pa
A}S^{(0)}=0\non
\del S = 0 &=& \eps\left( \del^{(1)}\phi\frac{\pa}{\pa
\phi}S^{(0)}+ \del^{(0)}A\frac{\pa}{\pa
A}S^{(1)}\right) \non
&+& \eps^2\left( \del^{(2)}\phi\frac{\pa}{\pa
\phi}S^{(0)}+ \del^{(1)}\phi\frac{\pa}{\pa
\phi}S^{(1)}+\del^{(0)}A\frac{\pa}{\pa
A}S^{(2)}+ \del^{(1)}A\frac{\pa}{\pa
A}S^{(1)}\right) + \ldots
\label{noet}
\end{eqnarray}
The linear term in $\eps$ is just the coupling of the gauge field to
the conserved current. The higher order terms are familiar from 
Yang-Mills theory; there $\del^{(2)}\phi =0$ vanishes and $\del^{(1)}A$
together with $S^{(2)}$ completes all derivatives to covariant
ones. There is no need to go to higher order in this case. For the
particle action in section two, one has to go to all orders, but the
net effect is just to complete the derivative in the variation of
$h_{ij}$ to a covariant one.

If we apply the Noether method to the $N$ D0-brane action, we find 
after a long and only partially 
illuminating calculation that one must guess higher
order terms {\em both} in the variation of the fields and the
action. These higher order terms should be of
the form 
\begin{eqnarray}
\label{noett}
\del^{(2)}X \sim \pa h \xi [[X,X],X] \non
S^{(2)} \sim h_{mn}h_{ij}[X,\dot{X}]^2
\end{eqnarray}
where $S^{(2)}$ is symmetric in $(mn) \leftrightarrow (ij)$.

Since the extra variation of $X$ depends explicitly on the
metric, this explains why it is not possible to lift
{\bf diff} to {\bf DIFF}. In fact, the Noether procedure
shows something even stronger, it shows that eq. (\ref{eq:3bagain})
cannot be achieved unless we allow an explicit dependence
on the metric in $X'(X)$. In other words, we will need
to replace eq. (\ref{eq:3b}) by the requirement
\be \label{eq:3c}
 S[g,X]=S[g'(g),X'(g,X)] .
\ee
In order to have the right point particle limit, the $g$ dependence
in $X'(g,X)$ should be only through commutator terms and
drop out when $X$ is taken diagonal. The fact that diffeomorphisms
depend explicitly on the background metric is reminiscent of
the Seiberg-Witten map in non-commutative gauge theory \cite{seibergwitten}
where the gauge parameter of the commutative theory depends
non-trivially on the non-commutative gauge field. Here
the matrix-valued gauge transformation depends non-trivially
on the gauge field for space-time gauge transformations, which is
the space-time metric.

We will take (\ref{eq:3c}) as our basic requirement of diffeomorphism
invariance. In the next section, we will try to put it in more
manageable form, and study its solutions.

\section{Matrix-valued diffeomorphisms from open-strings}

Eq. (\ref{eq:3c}) shows that for diffeomorphism
  invariance for multiple D-branes, 
we need to introduce a new type of symmetry. It is not
a priori clear what this symmetry is, and how we can
  use it the determine the form
  of the effective action. This effective action $S[g,X]$ 
for multiple D-branes coupled
to a background metric, can, to lowest order in the string coupling,
in principle be derived by
considering open string disc amplitudes with arbitrarily many
open string vertex operators attached to the boundary and 
arbitrarily many graviton vertex operators inserted in the
interior. As string theory is diffeomorphism invariant, the
  action derived in perturbation theory will automatically satisfy
  (\ref{eq:3c}). We can thus study properties of the transformations
(\ref{eq:3c}) from string perturbation theory,
  even though the symmetry (\ref{eq:3c}) is not manifest.

Amplitudes with one graviton vertex
operator have been computed in \cite{Okawa:2001if} and yield
the symmetrized trace prescription previously found by
\cite{dec,apr}. Eq. (\ref{noett}) shows that new terms that are not symmetrically
ordered appear once we consider two graviton vertex operators.
We will not attempt to compute this or any other amplitude here, but we will 
only examine their general structure. We
  will then use this information to constrain the form of the action and
  these constraints will allow us to find one which obeys eq. (\ref{eq:3c}).

 The effective action only follows indirectly from
  string-amplitudes and one may wonder why we do not use
  beta-functions to reconstruct the action. It is important to realize that
  standard beta-function methods do not work, because 
  we cannot turn on an 
arbitrary background for the matrix-valued transversal
scalar fields.\footnote{An attempt to perform such
a calculation was made in \cite{Dorn}, but the answer
found there is completely symmetrically ordered, and
as we already saw, this is not consistent with diffeomorphism
invariance.} That is because the off-diagonal elements
of the transversal matrix-valued scalar fields correspond
to open strings stretched between D-branes and these
are generically massive. Although one can in principle try to 
directly compute effective actions for massive string degrees of freedom,
see e.g. \cite{massive}, it is not clear how to do that in the
present context. We therefore have to resort to the calculation
of correlation functions.

This raises in consequence the crucial question whether a local
well-defined action for the transversal matrix-valued fields actually exists,
given that the only quantity  unambiguously available from string theory is
the S-matrix. If we believe that there is a local formulation
of string field theory, the action for $N$ D-branes
can be computed by integrating out massive string degrees of
freedom. One expects a local action as long as the
masses of open strings stretched between the branes is much
smaller than the masses of all other massive open string degrees
of freedom. This implies that the expectation value of the
scalar fields (except for the $U(1)$ part)
has to be much smaller than $l_s$. In addition, the momenta of
these fields (along the brane) have to be much smaller than
$1/l_s$ so that we can neglect higher derivative terms,
and string loop effects are suppressed as long as 
the string coupling $g_s \ll 1$. Our results will be applicable
in this regime only. Indeed, provided the graviton momenta are chosen
to scale in an appropriate way, it appears possible to find a consistent
$\alp\pr \rar 0$ limit for multiple graviton scattering, just as one is able
to do for a single graviton or other closed string modes
\cite{Okawa:2001sh,Okawa:2001if,Liu:2001ps,Das:2001ur,leigh}.

\bigskip

So, what general properties of the effective action can we deduce from
string perturbation theory, without computing any amplitudes?
The correlation functions that are relevant are correlation
functions of vertex operators for the transversal scalar fields and gravitons
\begin{equation} \label{vop}
V_{\mu,\alpha\beta}(k)\pa_{n}Y^{\mu}e^{ikY}(t)~~~,
~~~~~~h_{\mu\nu}(p)\pa Y^{\mu}\bar{\pa}\bar{Y}^{\nu}e^{ipY}(z,\bar{z})
\end{equation}
where $Y^{\mu}$ are the world-sheet fields, and $t$ is a point on
the boundary of the open-string worldsheet. 
The boundary conditions on $Y^{\mu}$ are that $Y^i(0)=Y^i(\pi)=\bar{x}^i$
for the Dirichlet directions $i=1,\ldots,9$, and Neumann boundary
conditions on $Y^0$, 
and we have explicitly written out the Chan-Paton indices $\alpha,\beta$.

The effective action that one obtains from string theory is
therefore of the form
\be \label{seff1}
S=S[h_{ij}(\bar{x}),\tilde{X}^i]_{\eta_{ij},\bar{x}^i}
\ee
where $\eta_{ij}$ and $\bar{x}^i$ label the background configuration,
and $h_{ij}$ and $\tilde{X}^i=(\eta+h)^{ij}V_j$  
the fluctuations in the metric
and transversal scalars respectively. 

It is the action (\ref{seff1})
that we want to relate to an action $S[g(X),X]$. That such an
relation should exist follows from the fact that turning
on a zero-momentum vertex operator (\ref{vop}) 
proportional to the identity matrix $\delta_{\alpha\beta}$,
corresponds to an infinitesimal change of the boundary condition
$Y^i(\sig=0,\pi)=\bar{x}^i$. 
Similarly, we can perform a finite shift of $\bar{x}^i$
by turning on a suitable condensate of the vertex operators
(\ref{vop}). This implies that the $\bar{x}^i$ dependence of
(\ref{seff1}) can be absorbed completely in $\tilde{X}$, and
the final result is an action of the form
$S[g(X),X]$. This remark is crucial in our quest for a
diffeomorphism invariant action (\ref{eq:3c}). 
The polarization tensors $V_{i,\alp\bet}(k)$ transform under a
space-time diffeomorphism $\del \bar{x}^i
=\xi(\bar{x})^i$ as a vector. Therefore, the effective
action obtained from string theory, $S[h_{ij}(\bar{x}),
\tilde{X}^i]_{\eta_{ij},\bar{x}^i}=S[g(X),X]$
will be invariant under diffeomorphisms.  Naturally 
$\eta_{ij} + h_{ij}$ transforms as a metric, and we expect $\tilde{X}^i$
to transform as a vector.  A priori though, 
there can be higher order corrections
to the transformation rule of $\tilde{X}^i$ due to contact terms. 
Indeed, the argument at the
beginning of this paragraph showed that $\tilde{X}$ is naturally a
difference between two points, rather than a vector. 
However, if we employ a covariant background field expansion (see
e.g. \cite{freed,Mukhi:1986vy})
to perform world-sheet
calculations, all quantities that appear in the calculation
will be tensors. Therefore we expect that there exists a choice of 
fields $\tilde{X}^i$ such that there are no higher order corrections
to the transformation rule of $\tilde{X}^i$.

This appears to imply that the diffeomorphism constraint is  
not very restrictive; {\em any} action whose
terms are contractions of space-time tensors constructed out
of $g_{ij}(\bar{x})$ with products of vector-fields $\tilde{X}^i$
is diffeomorphism invariant. However, since the final
action can be written in the form $S[g(X),X]$, there should
also be symmetries of the action that shift $\bar{x}$ and
act non-trivially on $\tilde{X}$, but that preserve the ``coordinate''
  $X$ and
therefore also the action $S[g(X),X]$. The existence of such symmetries
will severely constrain the action. 
The common setting to study
these symmetries is the covariant background field expansion. 
We will first discuss the closely related
Riemann normal coordinate system; and return to the issue
of general coordinates in section~5.2. 

\subsection{Normal coordinates}
\subsubsection{Particle}
As we explained above, $\bar{x}$ can be absorbed in the
field $\tilde{X}$, and the final action depends only on a 
field $X$ which is of the form
\be X = \bar{x} + \tilde{X} + {\cal O}(\tilde{X}^2) .
\label{exp1}
\ee
The higher order terms can in principle be computed using
higher order correlation functions in string theory, but
we will not attempt to do such a calculation here.
Similar higher order terms appear
in the case of a single D-brane,  for which  
the covariant non-linear 
background field expansion,
\cite{freed}
\begin{equation}
x^i =  \bar{x}^i + \tilde{x}^i - \sum_{n=2}^{\infty}\ove{n!}
\Gam^i_{j_1\ldots
j_n}(\bar{x})\tilde{x}^{j_1}\ldots \tilde{x}^{j_n}~.
\label{exp}
\end{equation}
guarantees that $\tilde{x}^i$ transforms as a vector under
diffeomorphisms in the backgroundfield $\del\bar{x}^i=\xi^i(\bar{x})$.
The pseudo-tensors $\Gam^i_{j_1\ldots
j_n}(\bar{x})$ equal $\Gam^i_{j_1\ldots
j_n}(\bar{x}) \equiv
\bar{\nabla}_{(j_1}\ldots\bar{\nabla}_{j_{n-2}}\Gam^i_{j_{n-1}j_n)}(\bar{x})$
where $\bar{\nabla}_{j_1}$ acts only on the lower
indices. The action
$S_{\rm particle}[g_{ij}(x(\bar{x},\tilde{x})),x^i(\bar{x},\tilde{x})]$
expressed in 
terms of $\bar{x}^i$ 
and $\tilde{x}^i$ will contain only proper tensors evaluated at
$\bar{x}^i$ contracted with the vectors $\tilde{x}^i$,
$\dot{\bar{x}}$.
For an easy algorithm to generate this expansion,
see \cite{Mukhi:1986vy}. The proper 
tensors must be built from the metric or derivatives thereof, so
that (suppressing time derivatives):
\begin{equation}
S_{\rm particle}[g_{ij}(x(\bar{x},\tilde{x})),x^i(\bar{x},\tilde{x})] =
S[g_{ij}(\bar{x}), D_{i_1}\ldots D_{i_n}R_{jklm}(\bar{x}), 
\tilde{x}^i] . \label{aux9}
\end{equation}
Again, one might have expected the full non-abelian action to be some
ordered version of (\ref{aux9}), where $\tilde{x}$  is replaced 
by $\tilde{X}$. If we assume that $\tilde{X}$ transforms as
a vector, then under space-time diffeomorphisms the action obtained
by ordering (\ref{aux9}) will be diffeomorphism invariant for {\em
  any} 
ordering.

\medskip
Normal coordinates is the special coordinate system where all the
non-linear terms in (\ref{exp}) vanish. In this coordinate system
covariant derivatives at $\bar{x}^i$ thus equal regular
derivatives. Therefore,
 in normal coordinates the action (\ref{seff1}) 
will have an expansion 
in proper space-time tensors,
similar to (\ref{aux9}).
As (\ref{exp})
is obtained from a perturbative solution to the geodesic
equation/field equation
\begin{equation}
  \label{eq:5}
  \ddot{x}^i+\Gam^i_{~jk}\dot{x}^j\dot{x}^k=0
\end{equation}
this shows that in normal coordinates, geodesics through the origin
(the point $\bar{x}$) are straight lines. Equivalently, in normal
coordinates $x^i(\tau)=\tau y^i$ is a solution to the field equation.

Vice versa, the action in normal coordinates can be used  
to obtain the covariant
action \cite{freed}. So given
the action in normal coordinates, we can easily construct the action
in any coordinate system. In the next section, we will discuss
the implications of
diffeomorphism invariance for the action in normal coordinates,
but first we discuss in some more detail the case when $X$ is matrix-valued.

\subsubsection{Matrices}

The action for matrix-valued fields is a special case of the
$dN^2$-dimensional point particle action 
\be S  \sim \int d\tau G_{IJ} \dot{X}^I \dot{X}^J 
\label{aux8}
\ee
introduced earlier,
where $I$ denotes the set of indices $i\alp\bet$. 

Any action of type (\ref{aux8}) yields a geodesic equation and
we can  use this to perform a covariant background field expansion
for (\ref{aux8}), exactly as we did above for a point particle
in $d$ dimensions. In particular, we will  specialize to 
\be
\label{P}
X = \bar{x} \one + \tilde{X} + \ldots
\ee
Substituting this expansion back in the action (\ref{aux8}) we
will encounter tensors evaluated at $\bar{x}\one$, with 
indices $I,J,K,\ldots$. They are tensors under  $d$-dimensional
 {\em matrix}
diffeomorphisms of the background coordinate $\bar{x}^i$.
 The group of matrix diffeomorphisms, evaluated on a diagonal
matrix, gives back the group of ordinary space-time diffeomorphisms.
Even more is true: take a matrix-valued diffeomorphism
compatible with the single trace structure, $X^I \rightarrow F(X)^I$,
then
\be \left.
\frac{\partial F(X)^{i\alpha\beta} }{\partial X^{j\gamma\delta} }
\right|_{X=\bar{x}\one} =
\left( \frac{\partial F(\bar{x})^i }{\partial \bar{x}^j }\right)
\delta^{\alpha}_{\gamma} \delta^{\beta}_{\delta} .
\label{vardiff}
\ee

We now define matrix normal coordinates as that choice of coordinates
where (\ref{P}) is linear. Due to the $dN^2$-dimensional origin it is
clear that matrix normal coordinates are equivalent to
$X^i(\tau)=\tau Y^i$ being a solution to the field equation.
Furthermore, in
matrix normal coordinates the action contains only
$dN^2$-dimensional proper tensors, namely the metric, the Riemann tensor and
its covariant derivatives, and is of the form (\ref{aux9}) with
$dN^2$-dimensional indices $I,J,K,\ldots$. We already argued
that these $dN^2$-dimensional tensors are constructed out of
$d$-dimensional space-time tensors in section~2.  Equation
(\ref{vardiff}) shows that this is also a sufficient condition
for the matrix-valued action to be invariant under matrix-valued
diffeomorphisms of the background field.
It is the diffeomorphisms that are not purely diffeomorphisms
of the background field that will significantly constrain
the action expanded in space-time tensors, and it is these
diffeomorphisms that we discuss next.

\subsection{Diffeomorphism in normal coordinates: \\
Base point independence}

We saw that the choice of normal coordinates was quite convenient,
since all resulting structures can be expressed  conventional
space-time tensors. To understand the consequences of diffeomorphism
invariance in normal coordinates, we reconsider the case of the
point particle. 

If the action (in arbitrary coordinates) is truly general coordinate
invariant, 
then we could have also chosen to expand around a second set of normal
coordinates $\tilde{z}^i$ centered around a {\em different} point
$\bar{z}$. The final action $S[g_{ij}(\bar{z}), D_{i_1}\ldots
D_{i_n}R_{jklm}(\bar{z}) , 
\tilde{z}^i]$ should be {\em identical} in form to $S[g_{ij}(\bar{x}),
D_{i_1}\ldots D_{i_n}R_{jklm}(\bar{x}),  
\tilde{x}^i]$ under the formal replacement $\bar{x} \leftrightarrow
\bar{z}$, $\tilde{x} \leftrightarrow
\tilde{z}$. 

General coordinate invariance also means that the action is invariant
if we perform the
coordinate transformation from the normal coordinate system 
$\tilde{x}^i$ around $\bar{x}^i$ to the normal coordinate system 
$\tilde{z}^i$ around $\bar{z}^i$. 

To obtain from an arbitrary
coordinate system  --- including another system of normal coordinates
--- a system of Riemann normal coordinates around
the base point $\bar{z}$, we consider geodesics of the form
\be
x^i(\tau) = \bar{z}^i + \tau \tilde{z}^i + {\cal O}(\tau^2),
\ee
so that $x^i(0)=\bar{z}^i$ and $\dot{x}^i(0)=\tilde{z}^i$.
The higher order terms are uniquely fixed by the geodesic equation
\begin{equation}
\ddot{x}^i + \Gam^i_{~jk}\dot{x}^j \dot{x}^k =0 .
\end{equation}
By definition, $x^i(\tau=1)$ is the expression for $x^i$ in
terms of Riemann normal coordinates $\tilde{z}^i$
around $\bar{z}^i$. In other
words, we flow with unit time along a geodesic that starts
at $\bar{z}^i$ and has initial velocity $\tilde{z}^i$. 

Specifically we could consider the second normal
coordinate system $\tilde{z}$ to be centered infinitesimally close to
$\tilde{x}$: $\bar{x}^i-\bar{z}^i = \eps^i \ll 1$. Now since all
parameters in the action for $\tilde{x}^i$ are proper tensors at
$\bar{x}^i$  the total effect of the coordinate transformation is
  that they are just parallel transported by an infinitesimal
amount $\eps^i$. Explicitly this means that
all terms of order $\eps$ or higher
should cancel order by order in 
\begin{eqnarray}
&&S[g_{ij}(\bar{x}), D_{i_1}\ldots
D_{i_n}R_{jklm}(\bar{x}) , 
\tilde{x}^i] \non
&&= S[g_{ij}(\bar{x}(\bar{z}+\eps)), D_{i_1}\ldots
D_{i_n}R_{jklm}(\bar{x}(\bar{z}+\eps)) , 
\tilde{x}^i(\tilde{z} + \eps(\tilde{z})+\cO(\eps^2))] \non
&&= 
S[g_{ij}(\bar{z})+\eps^kD_kg_{ij}(\bar{z}), D_{i_1}\ldots
D_{i_n}R_{jklm}(\bar{z})+\eps^kD_kD_{i_1}\ldots
D_{i_n}R_{jklm}(\bar{z}) , 
\tilde{z}^i + \eps(\tilde{z}); \cO(\eps^2)] \non
&&\equiv
S[g_{ij}(\bar{z}), D_{i_1}\ldots
D_{i_n}R_{jklm}(\bar{z}) , 
\tilde{z}^i]
\end{eqnarray}

An identical statement is true for $\tilde{X}^i$ and
$\tilde{Z}^i$ matrix-valued. In that case, the relevant 
tensors are tensors in the $dN^2$ dimensional space
described by the matrix-valued coordinates. In other
words, all indices should be replaced by multi-indices
$I \equiv i;\alpha\beta$. Now it is crucial that in the
matrix-valued case all tensors in the $dN^2$ dimensional
space have to be constructed out of ordinary space-time
tensors. 
Once we express the tensors in the large
$dN^2$ dimensional space by expressions in terms
of ordinary space-time tensors, the above statement
of base-point independence becomes a highly non-trivial
constraint, as we will see. 

\bigskip

To summarize, string theory gives us an action for multiple
D0 branes that depends on choice of base point $\bar{x}^i$,
a matrix-valued field $\tilde{X}^i$, and space-time tensors
evaluated at $\bar{x}^i$. This is achieved by using Riemann
normal coordinates in space-time. The action comes with a
fixed ordering, which is independent of $\bar{x}$. Without
loss of generality, we can always redefine $\tilde{X}^i$ so
that $\tilde{X}^i$ form a set of matrix Riemann normal
coordinates. (this is equivalent to requiring that $
\tilde{X}^i =  \tau Y^i$ is a solution of the field equations
for all constant $Y^i$). In that case, the resulting action should be 
base-point independent. This means that there should
exist field redefinitions of $\tilde{X}^i \rightarrow
\epsilon^i \one + \tilde{X}^i + {\cal O}(\tilde{X}^2)$
such that after the field redefinition the action looks identical,
including the ordering, except that the base point $\bar{x}$ 
has been replaced $\bar{x}+\epsilon$. The latter requirement
follows naturally from string theory: by condensing transversal
scalar fields one should be able to move a brane around in
space-time. World-sheet calculations for both the original and
final brane configuration are completely equivalent, and
should therefore give identical actions, including the ordering.

\subsection{Solving the constraints directly in the action}

Based on the discussion above, we will discuss two ways
to impose matrix normal coordinates and 
to solve the diffeomorphism constraint. In the next section,
we do this in terms of higher $dN^2$ dimensional geometry,
here directly using the action.

We first write down the most general action $S[g_{ij}(\bar{x}), D_{i_1}\ldots
D_{i_n}R_{jklm}(\bar{x}) , 
\tilde{X}^i]$ with arbitrary coefficients for all possible orderings
of the matrices $\tilde{X}^i$. There are always two matrices carrying
a proper time derivative, $d/d\tau$, 
and the location of these can differ depending on the 
ordering. The coefficients are
proper tensors evaluated at $\bar{x}$ and contracted in all possible
ways with a specific ordering of the matrices. Dropping the tilde on the
$\tilde{X}^i$ we have
\be
S = - \int d\tau \hlf \tr
(\dot{{X}}^i\dot{{X}}_i) + \sum_{p,n}
A^{(n)}_{i_1 \ldots i_p}
\tr(\dot{{X}}^{i_1}{X}^{i_2}\ldots
{X}^{i_{n-1}}\dot{{X}}^{i_n}{X}^{i_{n+1}}\ldots
{X}^{i_p})
\label{act}
\ee
From this action we derive the field equation
\begin{eqnarray}
\frac{\del S}{\del X^j} &=& \ddot{X}^j - \sum_{p,n,m\neq 1,n}
A^{(n)}_{i_1 \ldots i_{m-1} j i_{m+1} \ldots i_p}
(X^{i_{m+1}} \ldots X^{i_{n-1}} \dot{X}^{i_n}
X^{i_{n+1}} \ldots X^{i_p} \dot{{X}}^{i_1} \ldots
 X^{i_{m-1}} )
\non &&
+\sum_{p,n}
A^{(n)}_{j i_2 \ldots i_p}\frac{d}{d\tau}
({X}^{i_2}\ldots
{X}^{i_{n-1}}\dot{{X}}^{i_n}{X}^{i_{n+1}}\ldots
{X}^{i_p})
 \non &&
+ \sum_{p,n}
A^{(n)}_{i_1 \ldots i_{n-1}j i_{n+1} \ldots i_p}\frac{d}{d\tau}
({X}^{i_{n+1}}\ldots
{X}^{i_p}\dot{{X}}^{i_1}{X}^{i_2}\ldots
{X}^{i_{n-1}})
\label{eq2}
\end{eqnarray}
The first constraint is the requirement that the action $S$ we started
with in
eq.\rf{act} is in normal coordinates around $\bar{x}$. This is
equivalent to requiring that
$X^i(\tau) = \tau Y^i$ is a solution of the field equation \rf{eq2}.
Setting $\tau=1$ this imposes a constraint on the to be
determined coefficients $A^{(n)}$.
\begin{eqnarray}
0  &=& \sum_{p,n,m\neq 1,n}
A^{(n)}_{i_1 \ldots i_{m-1} j i_{m+1} \ldots i_p}
(Y^{i_{m+1}} \ldots Y^{i_{n-1}} {Y}^{i_n}
Y^{i_{n+1}} \ldots Y^{i_p} {{Y}}^{i_1} \ldots
 Y^{i_{m-1}} )
\non &&
- \sum_{p,n}
(p-2) A^{(n)}_{j i_2 \ldots i_p}
({Y}^{i_2}\ldots
{Y}^{i_{n-1}}{{Y}}^{i_n}{Y}^{i_{n+1}}\ldots
{Y}^{i_p})
 \non &&
- \sum_{p,n}
(p-2) A^{(n)}_{i_1 \ldots i_{n-1}j i_{n+1} \ldots i_p}
({Y}^{i_{n+1}}\ldots
{Y}^{i_p}{{Y}}^{i_1}{Y}^{i_2}\ldots
{Y}^{i_{n-1}})
\label{eq}
\end{eqnarray}

Next we find the coordinate transformation
to a nearby system of normal coordinates $Z^i$ around $\bar{z}
=\bar{x}+\eps$. Substituting $\dot{X}^i(0) = Z^i$ 
in the matrix-geodesic connecting 
$X^i$ with $\eps^i$ 
\begin{equation}
X^i(\tau) = \eps^i +\tau\dot{X}^i(0) + \frac{\tau^2}{2!} \ddot{X^i}(0) +\ldots
\end{equation}
and solving for higher derivatives of $X^i$ in terms of $Z^i$ and
$\eps^i$ using the field equation, we find the coordinate transformation
between $X^i$ and $Z^i$
\begin{equation}
X^i = \eps^i +Z^i + \sum_{p}
\Del^{i}_{j_1\ldots j_p}(\eps,A^{(n)}) Z^{j_1}\ldots Z^{j_p}
\label{trans}
\end{equation}
Note that the
imposition of the first constraint, that $X^i =\tau Y^i$ is a
solution to the field equation, means that in $\Del^{i}_{j_1\ldots
j_p}$ 
the zeroth order term
in $\eps^i$ vanishes. 
 Substituting \rf{trans} in the action, we keep the term linear in
  $\eps$, and denote it with $\del S(\eps,Z)$. This change in the
action should be identical to a shift in the base
point $\bar{x}^i \rightarrow {\bar z}^i + \eps^i$.
In normal coordinates this is achieved by parallel
transporting all tensors $A^{(n)}$ by
an infinitesimal amount $\eps$. Thus the condition
of base point independence is equivalent to the requirement that
\be
\del S(\eps,Z) = - \int d\tau
 \sum_{p,n}
\eps^k D_k A^{(n)}_{i_1 \ldots i_p}
\tr(\dot{{X}}^{i_1}{X}^{i_2}\ldots
{X}^{i_{n-1}}\dot{{X}}^{i_n}{X}^{i_{n+1}}\ldots
{X}^{i_p}) .
\label{act222}
\ee
In appendix~A.1 we work  out the details of
this procedure to order $X^4$.

In section 2 we saw that  the linearized coupling found by Taylor
and van Raamsdonk \cite{dec,apr} 
\begin{equation}
\label{ld:eq:1}
S= -\hlf\int\Tr \Xd^i\Xd_i + \hlf \int
\frac{1}{n!}\STr\left(\Xd^i\Xd^jX^{k_1}\ldots
  X^{k_n}\right)\pa_{k_1}\ldots\pa_{k_n}h_{ij} ~,
\end{equation}
is diffeomorphism invariant to first
order. It should therefore
satisfy the condition of base-point independence, and 
it is instructive to verify this. In terms of (matrix) 
Riemann normal coordinates, the linearized coupling equals
\begin{equation}
\label{ld:eq:2}
  \pa_{k_1}\ldots\pa_{k_n}h_{ij} =
\frac{\alpha_n}{n!}D_{(k_1}\ldots D_{k_{n-2}}
R_{k_{n-1}|(ij)|k_n)} + \cO(Riem^2) 
\end{equation}
where $\alpha_n =(n-1)/(n+1)$. 
In appendix A.2 we show that
this action is consistent with base-point independence. 
As one should expect, one can in fact derive 
explicit value of $\alp_n$ this way.
At higher orders the above
method becomes cumbersome and the 
 matrix geometry approach is more efficient.

With the use of the full Riemann tensor and symmetrized covariant
derivatives thereof, it may seem as if we have obtained more
information than the linearized results of Taylor and van
Raamsdonk. This is of course not true, as the above result only holds
in the normal coordinate frame, where the linearized expression for
the metric at $\pa\ldots\pa h(\bar{x})$ in (\ref{ld:eq:2})
exactly equals the symmetrized covariant
derivatives of the Riemann tensor.

\subsection{Solving the constraints in matrix geometry}

The action we start with is an action of the form
\be \label{act22}
S = \hlf \int d\tau g_{IJ}(X) \dot{X}^I \dot{X}^J 
\ee
where $I,J$ are $dN^2$ dimensional indices that
are of the form $i;\alpha\beta$ with $i=1,\ldots,d$ etc.
What we want is to express $G_{IJ}(X)$ in terms of $g_{ij}(x)$. 
Our strategy is
to first go to Riemann normal coordinates, in which
the metric has an expansion of the form 
\bea
g_{IJ}(X)&=&\delta_{IJ}
+{1 \over 3} R_{IKLJ}(\bar{x})X^{K}X^{L}
+{1 \over 6} {\nabla}_{M}R_{IKLJ}(\bar{x})
X^{K}X^{L}
X^{M}
\\ \nonumber
&& \hspace{-0.225in}
+{2\over 45}R_{IKL\lambda}(\bar{x})
R^{\lambda}_{\,\,\,MNJ}(\bar{x})
X^{K}X^{L}X^{M}X^{N}
+{1\over 20}
R_{IKLJ ;MN} (\bar{x}) X^K X^L
X^M X^N
+ O(X^5) .
\label{expan}
\eea
 Expressing $g_{IJ}(X)$ in terms of $g_{ij}(x)$, now amounts to
  expressing all curvature tensors and their
covariant derivatives $R_{IJKL}(\bar{x}), \nabla_M R_{IJKL}(\bar{x}),$ etc.
in terms of $R_{ijkl}(\bar{x}), \nabla_m R_{ijkl}(\bar{x}),$ etc.
In other
words, we will write down general expressions for
$R_{IJKL}$ and its derivatives in terms of $R_{ijkl}$ and its derivatives
times tensors containing all the Chan-Paton indices. 
For instance, we could try
\be
R_{IJKL} = R_{ijkl} 
 \delta_{\alp_2\bet_1
} \del_{\bet_2 \gam_1} \del_{\gam_2 \del_1} \del_{\del_2 
\alp_1}+\ldots
\ee
where $I=i;\alpha_1\alpha_2$, $J=j;\beta_1\beta_2$, etc.

We impose the following constraints on the tensors $R_{IJKL}(R_{ijkl},\ldots)$ and
their derivatives
\begin{itemize}
\item[({\em a})] The curvature tensor $R_{IJKL}$ has to have to usual symmetries,
meaning antisymmetric in the first and second pair of indices,
and symmetric under exchange of the first and second pair.
\item[({\em b})] It should have cyclic symmetry.
\item[({\em c})] Covariant derivatives $\nabla_I$ should obey the Bianchi identity.
\item[({\em d})] Multiple covariant derivatives should, when
anti-symmetrized, yield the usual rules, e.g.
$[\nabla_I,\nabla_J]R_{KLMN} = R_{IJK}{}^P R_{PLMN} + 
{\rm 3\,\,more\,\,terms}$.
\item[({\em e})] The 
$U(N)$ indices should be contracted in a single trace style, as the
action contains a single trace.
\item[({\em f})] To have the right behavior under change of basepoint, 
we require that 
\be
\delta^{\alp\bet} \nabla_{i\alp\bet}
({\rm anything}) = \nabla_i ({\rm anything})
\ee
\item[({\em g})] To first order in the curvature, the symmetrized
trace prescription, found in \cite{dec,apr} should emerge.
\item[({\em h})] It should have the right $U(1)$ limit for diagonal matrices.
\end{itemize}

Conditions ({\em a}) through ({\em d}) guarantee that the curvature
tensors and their derivatives all come from a single
metric $g_{IJ}(X)$. If we would just write down some
random tensors that would violate one of the conditions
({\em a}) through ({\em d}), they could never correspond to the
curvature tensors of some metric $g_{IJ}(X)$.

Condition ({\em e}) implies that the curvatures and
its covariant derivatives can be written as sums of
ordinary tensors with indices $i,j,k,l,\ldots$ times
tensors $\Delta_{\alp\bet\gam\del}$, where the latter
are defined as
\be
\Delta_{\alp\bet\gam\del} = \delta_{\alp_2\bet_1
} \del_{\bet_2 \gam_1} \del_{\gam_2 \del_1} \del_{\del_2 
\alp_1}
\label{Delta}
\ee
and similarly with more indices. The tensors $\Delta$ are 
cyclically invariant, and yield single trace expressions
when contracted with matrix-valued coordinates.

Condition ({\em f}) is crucial, and expresses the constraint of
base-point independence. In the way we have set things up,
it is guaranteed that under a change of base point all
tensors transform as $T \rightarrow \eps^i \delta^{\alpha_1\alpha_2}
\nabla_{i\alpha_1\alpha_2} T$. However, we want that
this is the same as changing the base point in all
ordinary tensors that appear in $T$, without affecting the
matrix structure. Therefore, we demand that
$\eps^i \delta^{\alpha_1\alpha_2}
\nabla_{i\alpha_1\alpha_2} T = \eps^i \nabla_i T$ for all $T$.

Finally, conditions ({\em g}) and ({\em h}) are obvious.

These are the ``axioms'' of the matrix geometry. Once
the curvature tensors have been specified, the action
in normal coordinates follows directly by applying the
expansion (\ref{expan}) to (\ref{act22}).

We have solved these equations for terms up to order $X^6$,
for which we need the curvature tensor and its first
and second derivatives.

There are 120 linearly independent terms containing 
two $\dot{X}$'s, four $X$'s, and a product of
two Riemann tensors with one index contracted, that one can write down.
We found that conditions ({\em a}) through ({\em h}) 
leave 32 of those undetermined.
The full result is very lengthy, and for illustrational purposes we
give a two-parameter family of solutions in appendix~B.
We postpone a discussion of this solution to section~5, and
conclude here by showing in some more detail how one determines
$R_{IJKL}$ in terms of $R_{ijkl}$.

\subsection{The matrix curvature tensor}

The most general form of the curvature that will give
a single trace answer is
\bea
R_{i\alp_1\alp_2 j\bet_1 \bet_2 k \gam_1 \gam_2 l
\del_1 \del_2} & = & 
T^1_{ijkl} \Delta_{\alp\bet\gam\del}  \non & & 
+ T^2_{ijkl} \Delta_{\alp\bet\del\gam}  \non & & 
+ T^3_{ijkl} \Delta_{\alp\gam\bet\del}  \non & & 
+ T^4_{ijkl} \Delta_{\alp\gam\del\bet}  \non & & 
+ T^5_{ijkl} \Delta_{\alp\del\bet\gam}  \non & & 
+ T^6_{ijkl} \Delta_{\alp\del\gam\bet} 
\eea
with $\Delta$ defined in (\ref{Delta}). 
The symmetries of the curvature tensor reduce this to
\bea
R_{i\alp_1\alp_2 j\bet_1 \bet_2 k \gam_1 \gam_2 l
\del_1 \del_2} & = & 
T^1_{ijkl} \Delta_{\alp\bet\gam\del}  \non & & 
- T^1_{ijlk} \Delta_{\alp\bet\del\gam}  \non & & 
T^3_{ijkl} \Delta_{\alp\gam\bet\del}  \non & & 
- T^1_{jikl} \Delta_{\alp\gam\del\bet}  \non & & 
+ T^3_{klij} \Delta_{\alp\del\bet\gam}  \non & & 
+ T^1_{jilk} \Delta_{\alp\del\gam\bet} 
\eea
where $T^1_{ijkl}=T^1_{klij}$, and $T^3$ is 
anti-symmetric in the first and second pair of
indices. Cyclic symmetry then implies
\bea
T^1_{ijkl} - T^1_{lijk} + T^3_{ljik} & = & 0 \non
-T^1_{ijlk} + T^3_{iljk} + T^1_{kijl} & = & 0 
\eea
Any of these two identities can be used
to write
\be
T^3_{ijkl} = T^1_{ikjl}- T^1_{kjli} 
\ee
where one can use the pair symmetry of $T^1$. 
Thus, everything is expressed in terms
of a single tensor $T^1$ that satisfies
\bea \label{q1}
T^1_{ijkl} - T^1_{klij} & = & 0 \non \label{q2}
T^1_{ikjl} - T^1_{kjli} + T^1_{jkil} - T^1_{kilj} & = & 0
\eea
where the second identity follows from the antisymmetry of
$T^3$. 
If $T^1$ can be expressed in terms of $R_{ijkl}$, it must
be of the form
\be T^1_{ijkl} = \lambda R_{ijkl} + \mu R_{ikjl}
\ee
Both (\ref{q1}) and (\ref{q2}) are satisfied for all
$\lambda$ and $\mu$. Thus symmetry dictates that the 
curvature is of the form
\bea
R_{i\alp_1\alp_2 j\bet_1 \bet_2 k \gam_1 \gam_2 l
\del_1 \del_2} & = & 
(\lambda R_{ijkl} + \mu R_{ikjl}) \nonumber
 \Delta_{\alp\bet\gam\del}  \non & & 
+(\lambda R_{ijkl} + \mu R_{ilkj} ) \nonumber
 \Delta_{\alp\bet\del\gam}  \non & & 
+(\lambda R_{ijkl} + 2 \mu R_{ijkl}) \nonumber
 \Delta_{\alp\gam\bet\del}  \non & & 
+(\lambda R_{ijkl} + \mu R_{ilkj} )  \nonumber
 \Delta_{\alp\gam\del\bet}  \non & & 
+ (\lambda R_{ijkl} + 2 \mu R_{ijkl}) \nonumber
 \Delta_{\alp\del\bet\gam}  \non & & 
+(\lambda R_{ijkl} + \mu R_{ikjl} )
 \Delta_{\alp\del\gam\bet}  \label{a1}
\eea
It is quite curious that a two-parameter family
of curvatures exists at this level. This can be
compared to the similar result derived in
appendix~A.1 where we implemented
diffeomorphism invariance directly in the
action. In the $U(1)$
limit we get $6(\lambda+\mu)R_{ijkl}$, and therefore
\be
6(\lambda+\mu) =1  .
\ee
To see what term one would get in the action, we contract with
\be
\dot{X}^{i\alp_1\alp_2} X^{ j\bet_1 \bet_2 } \dot{X}^{k \gam_1 \gam_2 } 
X^{l \del_1 \del_2} 
\ee
to obtain
\bea
\delta S 
 & = & 
(\lambda R_{ijkl} + \mu R_{ikjl})
 \tr(\dot{X}^i X^j \dot{X}^k X^l)   \non & & 
+(\lambda R_{ijkl} + \mu R_{ilkj} )
 \tr(\dot{X}^i X^j X^l \dot{X}^k )   \non & & 
+(\lambda R_{ijkl} + 2 \mu R_{ijkl})
 \tr(\dot{X}^i \dot{X}^k X^j X^l)   \non & & 
+(\lambda R_{ijkl} + \mu R_{ilkj} ) 
 \tr(\dot{X}^i \dot{X}^k X^l X^j)   \non & & 
+ (\lambda R_{ijkl} + 2 \mu R_{ijkl})
 \tr(\dot{X}^i X^l X^j \dot{X}^k)   \non & & 
+(\lambda R_{ijkl} + \mu R_{ikjl} )
 \tr(\dot{X}^i X^l \dot{X}^k X^j)   
\eea
We see that the $\mu$ terms do not contribute to the 
$\tr (\dot{X} X \dot{X} X) $ structure but do contribute
to the $\tr(\dot{X}\dot{X} X X)$ structure. Thus, in order
to get the symmetrized answer we need to take $\mu=0$.

To conclude, symmetries restrict the form to (\ref{a1}). 
If we in addition require the right form of the action,
i.e. completely symmetrized to leading order in $R$,
we need to take $\lambda=1/6$ and $\mu=0$, i.e.
\be R_{i\alp_1\alp_2 j\bet_1 \bet_2 k \gam_1 \gam_2 l
\del_1 \del_2} =
R_{ijkl} \Delta_{{\rm Sym} (\alp\bet\gam\del)} .
\ee
The Taylor-van Raamsdonk answer for the linearized
coupling is therefore not a consequence of diffeomorphism
invariance alone, but needs to be imposed by hand. 
This is confirmed by world-sheet calculations:
in the case of the bosonic string
it has been shown in \cite{Okawa:2001if} that the linearized
coupling to the metric is not completely symmetrically
ordered, but it should still be diffeomorphism invariant. 

A completely symmetrized answer is still a solution
for $\nabla R$, 
\be \nabla_{mE\eps_1\eps_2}
R_{i\alp_1\alp_2 j\bet_1 \bet_2 k \gam_1 \gam_2 l
\del_1 \del_2} =
 \nabla_m R_{ijkl} \Delta_{{\rm Sym}(
\eps\alp\bet\gam\del)} .
\ee
This structure breaks down for $\nabla\nabla R$, as one
can easily check that the completely symmetrized answer
does not give the right answer for $[\nabla,\nabla]R$
(point ({\em d}) in the criteria above).
Therefore, the action for $N$ D0 branes cannot
be the fully symmetrized answer of the action for
a single D0 brane. To get an idea of the complicated nature
of the action we give in appendix~B the action for
a special choice of parameters up to order $X^6$.

\section{Properties of the solution}

The action that we constructed explicitly up to order
$X^6$, and which is given in appendix~B, is quite ugly.
Even with 30 of the 32 free parameters removed, one is still 
left with a long expression. The result does exhibit 
properties of the general action that are worth observing.

One property is the fact that 
the action contains new contractions beyond 
the ones that are present 
in the $U(1)$ limit. For instance, in the $U(1)$ limit
the $R^2$ terms are always such that each $\dot{x}$ is
contracted with a separate curvature tensor. This is no
longer the case in the non-abelian action, which always
contains $R^2$ terms where two $\dot{X}$'s are contracted with
a single curvature tensor. This is clear from the
action given in appendix~B, and one can verify that it
remains true for arbitrary values of the 32 free parameters. 
Thus, the non-abelian action is not just a naive ordering
of the abelian one, but new terms are needed as well.

Once we extend the analysis to higher
order, at least some  
of the 32 free parameters will be fixed. For instance it is not hard
to show that the coefficient of the
term in the action proportional to $\Del S \sim
R_{ijkl}R^i_{mnp}{\rm Tr}({\rm Asym}(\dot{X}^jX^kX^l)
{\rm Asym}(\dot{X}^mX^nX^p))$
will only be determined by enforcing base-point independence at the
next order.
Ideally,
all of them would be fixed in this way, but this appears
unlikely. 
It is known that the
actions for D branes in the bosonic string and the superstring
are different once a non-trivial metric is introduced \cite{Okawa:2001if};
on the other hand both should be covariant. Therefore, there exists
more than one covariant non-abelian kinetic term. It is quite
likely that many covariant kinetic terms exist, that all differ
by commutator terms. It would be quite useful to have a
classification of all covariant kinetic terms, as this would
provide valuable insight in the structure of D-geometry.

\subsection{Reality of the action}

One issue we have not yet discussed at all is the reality of
the action. The fields $X$ are hermitian, and reality of the
action given in appendix~B will impose certain reality conditions
on the various free parameters. For example, reality of a combination
$a{\rm Tr}(X^i X^j X^k)+ b {\rm Tr}(X^k X^j X^i)$ requires that
$a$ and $b$ are each others complex conjugate. Once we determine
the reality conditions, one can in principle remove all
terms from the action that have imaginary coefficients in a
self-consistent way. It is possible that this will reduce the
number of free parameters somewhat, but we have not examined
this in detail. 

\subsection{Nonnormal coordinates}

Our analysis made crucial use of normal coordinates. We could
also have chosen to work in a completely arbitrary coordinate
system. In that case, the resulting action should be an 
expansion in $\tilde{X}^{i}$ whose coefficients are 
arbitrary multi-index objects built out of the metric and
its derivatives. Diffeomorphism invariance implies
that if we compute the action in two different
coordinate systems, and use the same regularization scheme
and method of computation for each, the resulting actions
should look exactly identical, including their ordering.
The only difference is in the explicit form of the metric that appears. 
Since the actions computed this way 
depend on arbitrary metrics and do not
rely on normal coordinates it may appear that covariance
of these actions can impose additional constraints.
This is not true, however. Each action can always be written
in terms of matrix Riemann normal coordinates. Since the
actions are identical, except for a change of metric, the
changes of coordinates will also be identical except
for the same change of metric. In Riemann normal coordinates,
we will therefore again end up with identical actions including
the ordering, except for the metric that appears. This
is precisely the structure that we worked with and we
see that it contains the same amount of information.

Conversely, suppose that we want to write down the action
for $N$ D0 branes in arbitrary coordinates. To accomplish
that, we first go to space-time normal coordinates and
construct the action as we explained. Next, we undo the 
change to space-time normal coordinates. This requires that
we find a matrix generalization of the map
from space-time normal coordinates to the coordinate
system we started out with. We can take any matrix
generalization that reduces in the $U(1)$ limit to the
required change of coordinates. Different choices simply
give actions that are related by a field redefinition and
therefore yield equivalent physics. If we agree on a 
specific ordering, e.g. completely symmetrized, and
use that for any change of normal coordinates to the original
coordinates, the resulting action in the original coordinates
will always look identical, including the ordering. The only
difference will be in the metric that appears. Thus, we
can construct an action for $N$ D0 branes in arbitrary coordinate
systems that has all the properties that we would expect
from string theory considerations. Different choices of orderings
  correspond to field redefinitions $\del X \sim [X,X]$ which involve
  commutators and vanish for diagonal matrices. 
  These are, of course, outside the scope of space-time
  diffeomorphisms.

\subsection{The potential, fermions and other contributions}

In this paper, we have so far focused exclusively on the kinetic
terms in the action. Of course, the complete non-abelian
DBI action has many more terms; it has a potential term,
fermionic terms, Wess-Zumino terms and higher order derivative
terms. Obviously, these terms all have to be separately
covariant. For those terms that only involve the 
transversal scalar fields $X$, the notion of covariance
is identical to that for the kinetic terms. In other
words, the terms should admit an expansion in terms
of space-time tensors multiplying expression in $\tilde{X}$,
and they should be base-point independent.  Suppose
that the kinetic term is given up to a certain order, then  we
  know the
transformations for $\tilde{X}$ that implement a change
of base-point up to that order. 
The same transformation should also yield
a change of base-point for all other terms in the
action, which are thereby severely constrained. 

For example, it is possible to write down a potential
in curved space 
in terms
of the matrix valued metric $G_{IJ}(X)$
that is completely covariant, and that reduces
to the usual potential in flat space.
To write the potential, we introduce a matrix version of
the vielbein by writing\footnote{Notice that these vielbeins
do not have simple $U(N)$ transformations rules. One can also
introduce twisted vielbeins via $G_{i\alpha\beta,j\gamma\delta}=
E^r_{i\alpha\delta} E^r_{j\gamma\beta}$, and which do transform
in the adjoint representation under $U(N)$. It is not clear
whether these twisted vielbeins play any role in matrix geometry.}
\be \label{viel}
G_{IJ}(X) = \sum_A E^A_{I^t} E^A_{J^t} , \qquad 
G^{IJ}E^A_{I^t} E^B_{J^t} = \delta^{AB},
\ee
where $A,B=1\ldots dN^2$. Here $I^t$ denotes $i\beta\alpha$ for
$I=i\alpha\beta$.
The base point independence of
the action $G_{IJ} \dot{X}^I \dot{X}^J$ implies a simple
transformation rule for $E^A_{I^t} \dot{X}^I$ under a change
of basepoint, namely it rotates by an $SO(dN^2)$ transformation.
Notice that $E^A_I \equiv E^A_{i\alpha\beta}$ is a matrix,
which we will denote by $E^A_i$. Thus it is ${\rm Tr}(E^A_i \dot{X}^i)$
that transforms nicely.
Now observe that taking the time
derivative acts as a derivation on the algebra of matrices
(i.e. it satisfies the Leibnitz rule), but so does the
operation $V:X \rightarrow [X,V]$ for fixed $V$. Therefore,
${\rm Tr}(E^A_i [X^i,V])={\rm Tr}([E^A_i,X^i]V)$ 
will also transform nicely under
a change of base-point, if we keep $V$ inert.
We claim that a covariant version of the potential is
\be \label{pott}
\frac{1}{2} {\rm Tr}([X^i,X^j]^2) \rightarrow \sum_{A,B}
\frac{1}{4} {\rm Tr} (
[E^A_i,X^i][E^B_j,X^j][E^A_k,X^k][E^B_l,X^l] ).
\ee
We just explained that $\tr [E^A_i,X^i]$ transforms under
a base-point changing transformation in a simply way,
via an $SO(dN^2)$ transformation. Eq. (\ref{pott}) is therefore 
automatically covariant. One can also verify
that in flat space, it reduces to the usual answer\footnote{Use that
  $G_{i\alp\bet,j\gam\del}|_{\mbox{\footnotesize flat}}=
  \eta_{ij}\del_{\alp\del}\del_{\bet\gam}$. See also the previous
  footnote. \label{fff2}}. Moreover, we  
  know the full linearized coupling of the potential term
to the metric \cite{dec,apr,Okawa:2001if}; it is given by a fully
symmetrized expression.  A somewhat
tedious calculation shows that this is correctly reproduced
by (\ref{pott}). This is strong evidence that eq. (\ref{pott}))
  is the curved-space version of the potential. One can try to
use a similar strategy to study the Ramond-Ramond couplings in
the action.

The use of vielbeins for the potential is encouraging, as its
introduction is a prerequisite for 
  the inclusion of fermions. Their 
correct treatment will demand a separate
  analysis. We
will get a structure where fermions and fields $\tilde{X}$ multiply
space-time tensors. However, the transformations that shift
the base-point may have to be modified. In the $U(1)$ limit,
such modifications are absent, but we do not know any reason
why they should be absent in the $U(N)$ limit as well. 

This also raises the question how world-volume supersymmetry
and $\kappa$-symmetry could be connected to our notion of covariance.
Both supersymmetry and $\kappa$-symmetry impose conditions on
the background fields, so we can no longer consider the most
general coupling. In particular, $\kappa$-symmetry typically 
requires the background fields to be on-shell. In such a
situation the constraint of covariance becomes less stringent.
On the other hand, to combine either supersymmetry or
kappa-symmetry with the non-abelian structure will significantly
constrain the possible actions that we can write down.
This idea was pursued for supersymmetry in \cite{douglas,kato}
and for $\kappa$-symmetry in \cite{Bergshoeff:2000ik}.

\subsection{The spectrum of quadratic fluctuations}

According the the axioms of D-geometry of Douglas,
the spectrum of quadratic fluctuations around a diagonal
matrix should agree with the geodesic lengths between
the space-time points appearing in the diagonal matrices.
This criterion is natural if we identify those diagonal
entries with the locations of the D-branes, and
the off-diagonal elements with the open strings stretched 
between them. We will now demonstrate that with the choice of
potential (\ref{pott}), the masses of the off-diagonal
fluctuations are indeed {\it exactly} given by the
geodesic distances between the separated branes.
For simplicity, we will assume that the action contains
terms with real coefficients only.

The masses of off-diagonal fluctuations are obtained by writing
the matrices $X$ as the sum of a diagonal matrix $\Delta$
with time-independent
diagonal entries $\lambda^i_{\alpha}$, plus an arbitrary off-diagonal
matrix $Y^i_{\alpha\beta}(t)$. We expand the action to second
order in $Y$ and to all orders in $\lambda^i_{\alpha}$, and from
the result we read off the masses.

We will work in the gauge $A_0=0$ and enforce the Gauss law
  explicitly. It states that\footnote{Recall that the action
  equals $S=\hlf\tr(E^A_i 
  D_0X^i)\tr(E^A_j D_0X^j)$ (note that this does not violate the
  $U(N)$ single-trace structure).}
\be \label{gauss}
G_{i\alpha\beta, j\gamma\rho} \dot{X}^i_{\alpha\beta}
X^{p}_{\rho\delta} -
G_{i\alpha\beta, j\rho\delta} \dot{X}^i_{\alpha\beta}
X^{p}_{\gamma\rho} =0.\ee
The expansion of the kinetic term is easy, because the
$\lambda^i_{\alpha}$ are time-independent. We need the metric
$G_{i\alpha\beta,j\gamma\delta}$ evaluated on the diagonal
matrix with entries $\lambda^i_{\alpha}$. This is of the form (see
  footnote \ref{fff2})
\be
G_{i\alpha\beta,j\gamma\delta}(\Delta) \equiv
S_{ij\alpha\beta}(\Delta) 
\delta_{\alpha\delta} \delta_{\beta\gamma}.
\ee
Clearly, $S_{ij\alpha\beta}$ depends only on $\lambda^k_{\alpha}$
and $\lambda^k_{\beta}$. 
The fact that the metric is symmetric implies 
\be \label{sym1}
S_{ij\alpha\beta}=S_{ji\beta\alpha}. \ee
Moreover, we
assumed that the action is real with real coefficients,
which yields the additional symmetry
\be \label{sym2}
S_{ij\alpha\beta}=S_{ji\alpha\beta}. \ee

With this definition of $S$, the expansion of the kinetic term
is given by
\be \label{kinexp}
{\cal L}_{\rm kin} = S_{ij\alpha\beta} 
\dot{Y}^i_{\alpha\beta} 
\dot{Y}^j_{\beta\alpha}  .
\ee

Next we turn to the expansion of the potential term (\ref{pott}).
Although this is not obvious from the form (\ref{pott}), the
only contributions to second order in $Y$ arise when two of
the four explicit $X$'s that appear in (\ref{pott}) are 
diagonal, and two of them are off-diagonal. This is most easily seen
by writing (\ref{pott}) explicitly in terms of indices, and
by reexpressing the vielbeins in terms of the metric using 
(\ref{viel}). One finds that the
potential can also be expressed in terms of $S$ and is given by
\be \label{potexp}
{\cal L}_{\rm pot}  =   (S_{ik\alpha\beta} S_{jl\alpha\beta}
- S_{ij\alpha\beta} S_{kl\alpha\beta} )
(\lambda_{\alpha}^i - \lambda_{\beta}^i)
(\lambda_{\alpha}^j - \lambda_{\beta}^j)
Y^k_{\alpha\beta} Y^l_{\beta\alpha}  .
\ee

At the order we are working in, the Gauss law constraint translates
into
\be
\label{gauss2}
S_{ij\alpha\beta} (\lambda_{\alpha}^i - \lambda_{\beta}^i) 
\dot{Y}^j_{\alpha\beta} =0~~~\rar~~~S_{ij\alpha\beta} (\lambda_{\alpha}^i - \lambda_{\beta}^i) 
{Y}^j_{\alpha\beta} = const.\ee
This decouples one of the nine transverse scalars.
Because we want to expand around a solution, 
we put the constant in (\ref{gauss2}) equal to zero, which causes the
first term in eq. (\ref{potexp}) to vanish.

It is then
obvious from (\ref{kinexp}) and (\ref{potexp}) that the
masses of the eight remaining transversal scalars are
all identical and equal to
\be \label{masses}
M^2_{\alpha\beta} = S_{ij\alpha\beta}(\Delta)
(\lambda_{\alpha}^i - \lambda_{\beta}^i)
(\lambda_{\alpha}^j - \lambda_{\beta}^j) .
\ee
It is not yet clear that this expression equals the
geodesic distance squared. It is true that the geodesic distance
between two points can be written in the form (\ref{masses})
where $S_{ij\alpha\beta}$ has a good power series expansion in
$\lambda_{\alpha}^i,\lambda_{\beta}^i$. To prove that (\ref{masses})
is identical to the geodesic distance squared, the notion of
covariance we introduced in this paper is crucial. Because of
this covariance, we can without loss of generality put
one of the two points, say $\lambda_{\beta}^i$, equal to zero,
and at the same time we can use Riemann normal coordinates around zero.
The geodesic distance in Riemann normal coordinates between
$0$ and $\lambda_{\alpha}^i$ is simply given by
\be d^2(\lambda_{\alpha},0) = g_{ij}(0) \lambda_{\alpha}^i
\lambda_{\alpha}^j, \ee
so we are done if we can show that 
\be \label{verf}
S_{ij\alpha\beta}(\Delta) 
\lambda_{\alpha}^i \lambda_{\alpha}^j = 
g_{ij}(0) \lambda_{\alpha}^i
\lambda_{\alpha}^j 
\ee
in Riemann normal coordinates.
From the action we already know that $S_{ij\alpha\beta}=
g_{ij}(0) + {\cal O}(\lambda_{\alpha}^2)$, and all
higher order terms in $S$ are contractions of tensors
built out of the Riemann tensor with the $\lambda_{\alpha}^i$.
A simple counting argument shows that in 
$S_{ij\alpha\beta}(\Delta) 
\lambda_{\alpha}^i \lambda_{\alpha}^j$ every term, except the leading one
proportional to $g_{ij}(0)$, contains at least one Riemann tensor with
at least three indices contracted with $\lambda_{\alpha}^i$. This vanishes
by the Bianchi identity, and we conclude that (\ref{verf})
is indeed true. 

Because of the covariance and base-point independence of the metric,
the result that the masses of off-diagonal fluctuations equals
the geodesic length will remain valid in any coordinate system.
This leads to the interesting conclusion that the geodesic distance
in space-time can be directly read off from $G_{IJ}(X)$, evaluated
on diagonal $X$. There is no need to integrate a line element
along a geodesic. More precisely, if we write the kinetic term
in the form
\be 
\sum_t {\rm Tr}(P^t_{ij}(X) \dot{X}^i Q^t_{ji}(X) \dot{X}^j)
\ee
with some set $P^t_{ij},Q^t_{ji}$, then the geodesic distance
$d(x,y)$ satisfies
\be
d(x,y)^2 = \sum_t P^t_{ij}(x) Q^t_{ji}(y) (x-y)^i (x-y)^j.
\ee

As a final comment, we observe the the expansion of the
potential to second order also immediately implies that
vacuum manifold of the potential consists of $M^N/S_N$,
i.e. the moduli space of $N$ unordered points in $M$.
This was one of the axioms of D-geometry as put forward
in \cite{douglas}.

\subsection{Connection with non-commutative geometry}

The fact that for multiple D-branes, the transverse coordinates
are replaced by matrices, suggests a role for non-commutative
geometry in the action for multiple D-branes. In non-commutative
geometry, the space of functions is replaced by a non-commutative
algebra, and the obvious candidate here would be to consider
the algebra
\be
{\cal A} = C^{\infty}(M) \otimes M_N(C) .
\ee
This algebra does not yet carry any metric information. From the
representation theoretical point of view, it is very close to
the original algebra $C^{\infty}(M)$ (they are Morita equivalent).
Following Connes, the construction
of a Riemannian structure requires a spectral triple $({\cal A},{\cal H},D)$
which in addition to ${\cal A}$ also contains a Hilbert
space ${\cal H}$ and a self-adjoint operator $D$ obeying certain
properties \cite{connesbook}. It would be interesting to find 
triples $({\cal A},{\cal H},D)$ that describe in a natural
way metrics relevant for multiple D-branes, and that incorporate
the notion of covariance. The form of the action and
the potential suggest that the vielbein $E^A_I$ introduced
in (\ref{viel}) will play an important role in such a construction.

Another possible connection with non-commutative geometry 
is with the non-commutative geometry description of D-branes
in the presence of a $B$-field in space-time (or a magnetic
field strength on the brane) \cite{douglashull,connesdouglasschwarz}.
The point is that the structure of an abelian non-commutative
gauge theory is very similar to that of a non-abelian
commutative gauge theory. In both cases, fields no longer commute,
and the field strengths are non-linear. Moreover, non-commutative
gauge theories can be constructed starting from a non-abelian
commutative theory by expanding around suitable backgrounds and
taking $N\rightarrow \infty$ (see e.g. \cite{seiberg0008013} and
references therein). The latter connection shows it should be
possible to relate the coupling of non-commutative gauge theories
to gravity to the coupling of non-abelian D-brane actions to
gravity. This was indeed the approach taken in \cite{Das:2001ur}
where the stress-tensor of
non-commutative gauge theories was derived in this way. It would be very
interesting to see whether a similar connection can be made
at higher order. The following remarks suggestive that
such connections may exist.
\begin{itemize}
\item[1]
There is a close relation between the world-sheet calculations
involving gravitons for non-commutative gauge theories and
D-branes \cite{Liu:2001ps,Okawa:2001if}. Both
lead to a result where certain operators are smeared along
some kind of Wilson line. This structure becomes more complicated
when more than one graviton vertex operator is 
involved\footnote{We would like to thank H. Ooguri for a discussion
of this point.}, and the precise geometry underlying such
calculations has not been uncovered.
\item[2]
Non-commutative spaces are constructed in deformation quantization 
from commutative spaces equipped with a closed two-form, and
this is also how they arise in string theory. The diffeomorphisms
of the commutative space that preserve the two-form become
gauge transformations of the non-commutative space (they
are just canonical transformations). Thus, there should also
be a relation between the coupling of a gauge field in non-commutative
gauge theory and the coupling of the graviton in non abelian
gauge theory. 
\item[3]
There is a one-to-one correspondence between single trace expressions
one can write down in terms of matrix valued coordinates $X^i$, and
expressions involving ordinary coordinates and a closed two-form
$B^{IJ}$.  Open string amplitudes depend only on the combination
$\cF=B+F$ and T-duality maps $F \leftrightarrow [X,X]$. This map
extends to 
\bea
X^i X^j + X^j X^i & \leftrightarrow & 2x^i x^j \nonumber \\
\mbox{} [X^i,X^j] & \leftrightarrow & B^{ij} \nonumber \\
\mbox{}[X^k,[X^i,X^j]] & \leftrightarrow & B^{kl} \partial_l B^{ij}
\eea
etcetera.\footnote{Due to the Jacobi relation, the $B$-field must obey
  certain constraints. These are equivalent to the integrability
  constraints in deformation quantization \cite{ksevich}.} 
This suggests there should be a relation (some kind
of Seiberg-Witten map) between a single D brane in a transversal
B-field, and multiple D-branes without a transversal B-field.
\end{itemize}

\section{Discussion and conclusion}

In this paper we discussed the notion of covariance for 
multiple D-branes in transversal curved space. We used
this notion to constrain the action and studied the kinetic
term explicitly up to sixth order in the fields. We also
gave an explicit form of the potential in terms of the
kinetic term, and showed that the resulting action satisfies
all axioms of D-geometry. 

We have uncovered a glimpse of an intricate geometrical
structure that encodes the behavior of multiple D-branes
in curved space. The precise mathematical structure underlying
this geometry, and the corresponding stringy fuzziness of
space-time, are still waiting to be uncovered.

Nevertheless, there are several directions in which the results
here can be extended. A covariant formulation of the coupling
to other closed string background fields, and an investigation
of analogs of the Myers effect in curved backgrounds are two
such issues. We would also like to know the behavior of D-branes
in black hole backgrounds, where we expect the deviations
from classical geometry to become particularly relevant.

\section{Acknowledgements}

We would like to thank K.~Bardakci, M.~Douglas,
R.~Dijkgraaf, H.~Ooguri, W.~Taylor and A.~Tseytlin
for discussions.

This work is also supported in part by NSF grant PHY-9907949
at ITP, Santa Barbara.  JdB.
is happy to thank the Institute for Theoretical Physics at
UCSB for hospitality during the penultimate phase of this project. KS
thanks the CNYITP at Stony Brook for hospitality. 

\appendix

\section{Diffeomorphism invariance in the action: explicit calculations}

In this appendix we give some explicit details of calculations
where we impose normal coordinates and diffeomorphism invariance
directly in the action. In section~A.1 we construct the action up
to order $X^4$, and in section~A.2 we give the linearized coupling
to the graviton to all orders.

\subsection{A calculation to order $X^4$}

We see from (\ref{act222}) that the constraint at
order $p$ in $Z^i$ requires one to start at order $p+1$ in the action
\rf{act}. The first non-trivial constraint will arise at order
three. The most general action with arbitrary coefficients for all
inequivalent orderings of the matrices $X^i$ to order four is 
\begin{eqnarray}
S &=& -{T_0} \int d\tau \hlf\tr
(\dot{{X}}^i\dot{{X}}_i) + 
A^{(3)}_{i_1i_2 i_3}
\tr(\dot{{X}}^{i_1}\dot{X}^{i_2}{X}^{i_3}) \non &&+ 
A^{(4);(1)}_{i_1i_2 i_3i_4}
\tr(\dot{{X}}^{i_1}\dot{X}^{i_2}{X}^{i_3}X^{i_4})+ 
A^{(4);(2)}_{i_1i_2 i_3i_4}
\tr(\dot{{X}}^{i_1}X^{i_2}\dot{X}^{i_3}{X}^{i_4}) + \ldots
\end{eqnarray}
Up to a factor of $-T_0$, this action has as  field equation 
\begin{eqnarray}
\frac{\del }{\del X^j}S &=& - \ddot{X}_j  -  A^{(3)}_{ji_2
i_3}\frac{d}{d\tau}(\dot{X}^{i_2}{X}^{i_3}) - A^{(3)}_{i_1j
i_3}\frac{d}{d\tau}({{X}}^{i_3}\dot{X}^{i_1}) 
+A^{(3)}_{i_1i_2 j}\dot{{X}}^{i_1}\dot{X}^{i_2} \non
&&\hspace{-.7in}
-A^{(4);(1)}_{ji_2 i_3i_4}\frac{d}{d\tau}(\dot{X}^{i_2}{X}^{i_3}X^{i_4})
-A^{(4);(1)}_{i_1j i_3i_4}\frac{d}{d\tau}
({X}^{i_3}{X}^{i_4}\dot{X}^{i_1})+
A^{(4);(1)}_{i_1i_2 i_3i_4}
({X}^{i_4}\dot{X}^{i_1}\dot{X}^{i_2}\del^{i_3}_j+\dot{X}^{i_1}\dot{X}^{i_2}
{X}^{i_3}\del^{i_4}_j) \non \nonumber
&& \hspace{-.7in}
-A^{(4);(2)}_{ji_2 i_3i_4}\frac{d}{d\tau}({X}^{i_2}\dot{X}^{i_3}X^{i_4})
-A^{(4);(2)}_{i_1 i_2ji_4}\frac{d}{d\tau}
({X}^{i_4}\dot{X}^{i_1}{X}^{i_2})+
A^{(4);(2)}_{i_1i_2 i_3i_4}
(\dot{X}^{i_3}{X}^{i_4}\dot{X}^{i_1}\del^{i_2}_j+\dot{X}^{i_1}{X}^{i_2}
\dot{X}^{i_3}\del^{i_4}_j)
\end{eqnarray}
The condition that $S$ is in fact in normal coordinates requires
that each of the combinations 
\begin{eqnarray}
0 &=&
\left[A^{(3)}_{ji_2i_3}+A^{(3)}_{i_3ji_2}-A^{(3)}_{i_2i_3j}\right]Y^{i_2}Y^{i_3}
\non
&&+ \left[2A^{(4);(1)}_{ji_2 i_3i_4} +2A^{(4);(1)}_{i_4ji_2
i_3}-A^{(4);(1)}_{i_3i_4ji_2}-A^{(4);(1)}_{i_2 i_3i_4j} \right. \non
&& \left.+2A^{(4);(2)}_{ji_2 i_3i_4}+2A^{(4);(2)}_{i_3i_4ji_2}-A^{(4);(2)}_{i_4ji_2 i_3}-A^{(4);(2)}_{i_2 i_3i_4j}\right]Y^{i_2}Y^{i_3}Y^{i_4}
\label{cons1}
\end{eqnarray}
vanish.

Next we establish the form of the coordinate transformation to normal
coordinates $Z^i$ around the nearby basepoint $\eps^i$. This is given
by
\begin{equation}
X^i(\tau) = X^i(0) +\tau\dot{X}^i(0) + \frac{\tau^2}{2!} \ddot{X^i}(0)
+\ldots 
\label{eq3}
\end{equation}
for $\tau=1$ after substituting $X^i =\eps^i \cdot \one$ and
$\dot{X}^i(0) =Z^i$. To find an expression for $\ddot{X}^i(0)$ we use
the field equation. We find to order $\eps$ and order two in
$Z^i$ (recall that the zeroth order term in $\eps$ vanishes due to
\rf{cons1}) 
\begin{eqnarray}
\ddot{X}_j(0) &=&
-
\left[A^{(4);(1)}_{ji_2i_3i_4}+A^{(4);(1)}_{ji_2i_4i_3}+A^{(4);(1)}_{i_3ji_2i_4}
+A^{(4);(1)}_{i_3ji_4i_2}-A^{(4);(1)}_{i_2i_3ji_4}-A^{(4);(1)}_{i_2i_3i_4j}
\right.\non
&&\left. 
+A^{(4);(2)}_{ji_2i_3i_4}+A^{(4);(2)}_{ji_4i_2i_3}+A^{(4);(2)}_{i_3i_4ji_2}
+A^{(4);(2)}_{i_2i_3ji_4}-A^{(4);(2)}_{i_3ji_2i_4}-A^{(4);(2)}_{i_2i_4i_3j}\right]
Z^{i_2}Z^{i_3}\eps^{i_4} 
+\cO(\eps^2) 
\end{eqnarray}
Something special happens in the next step, when we substitute the
coordinate change \rf{eq3} into the action, but this is peculiar to
this order. It does not occur at higher orders. Making the
substitution we find that at order $Z^3$ all terms involving the
coefficients $A^{(4);(1)}_{i_1i_2i_3i_4}$ and $A^{(4);(2)}_{i_1i_2i_3i_4}$ cancel and the linear term in $\eps$ in
the action is just
\begin{eqnarray}
\del S|_{\eps} &=& \int d\tau \del A^{(3)}_{i_1i_2 i_3}
\tr(\dot{{Z}}^{i_1}\dot{Z}^{i_2}{X}^{i_3})
\non &=&
\int d\tau \del \eps^k D_k A^{(3)}_{i_1i_2 i_3}
\tr(\dot{{Z}}^{i_1}\dot{Z}^{i_2}{X}^{i_3})
\end{eqnarray}
The requirement of general coordinate invariance is thus that the
tensor $A^{(3)}_{i_1i_2 i_3}$ is covariantly constant. There is,
however, no proper tensor with three indices that one can construct from
the metric or derivatives thereof.\footnote{This is not quite true,
e.g. $ D^kR_{ki_2i_3i_4}$,
but consistency with the $U(1)$ limit allows only the use of a single
derivative on the metric.} Hence $A^{(3)}_{i_1i_2 i_3}$ 
in fact vanishes identically. At this order general coordinate
invariance imposes no constraint.  
 
We are thus left solely with the two constraints \rf{cons1}. The first
one is again trivially satisfied as there does not exist a proper
tensor with three indices. There exists one proper tensor built from
two derivatives of the metric and with four indices: the Riemann
tensor. The coefficients $A^{(4);(1)}_{i_1i_2i_3i_4}$ and
$A^{(4);(2)}_{i_1i_2i_3i_4}$ are thus proportional to a 
particular combination of Riemann tensors. The various identities that
the latter satisfies mean that there are only two independent combinations
and therefore
\begin{eqnarray}
A^{(4);(1)}_{i_1i_2i_3i_4} &=& a_1 R_{i_1i_2i_3i_4}+a_2R_{i_1i_3i_2i_4} \non
A^{(4);(1)}_{i_1i_2i_3i_4} &=& b_1 R_{i_1i_2i_3i_4}+b_2R_{i_1i_3i_2i_4}
\label{riem}
\end{eqnarray}
Our task has reduced to determining the coefficients $a_1,a_2$ and
$b_1,b_2$. Substituting \rf{riem} into the normal coordinate
constraint
\rf{cons1} we find that the coefficients must obey
\begin{equation}
((- 2a_1 - 2b_1)R_{i_1i_2i_3j}
          + (a_1
          + 4b_1
          + 6b_2)R_{i_1i_3i_2j}
          )Y^{i_1}Y^{i_2}Y^{i_3} = 0
\label{rnc}
\end{equation}

The Riemann normal coordinate and base-point independence constraints
thus leave at order three a two-parameter class of solutions. At order
four and five, the system remains 
under-determined. 
To fix this final ambiguity we will require
that the solution is consistent with the linearized result obtained in
\cite{dec}.  To analyze this, we
substitute the most general form for the order-four coefficients
\rf{riem} into the action to get 
\begin{eqnarray}
S^{(4)} &=& a_1 R_{i_1i_2i_3i_4}\tr(\dot{{X}}^{i_1}\dot{X}^{i_2}{X}^{i_3}X^{i_4})
    + a_2 R_{i_1i_3i_2i_4}\tr(\dot{{X}}^{i_1}\dot{X}^{i_2}{X}^{i_3}X^{i_4})
\non && + b_1 R_{i_1i_2i_3i_4}\tr(\dot{{X}}^{i_1}X^{i_2}\dot{X}^{i_3}{X}^{i_4})
        + b_2 R_{i_1i_3i_2i_4}\tr(\dot{{X}}^{i_1}X^{i_2}\dot{X}^{i_3}{X}^{i_4})
\label{lin4}
\end{eqnarray}
At this level there is a combination of Riemann tensors
which equates to a double derivative on $h_{ij}$ at the origin in normal
coordinates.\footnote{In normal
coordinates derivatives of the metric at the origin are related. Using  
\begin{equation}
\pa_{(\mu_1 \ldots} \pa_{\mu_{n-2}}\Gam^{\tau}_{~\mu_{n-1} \mu_n)} = 0
\end{equation}
one can show that
\begin{equation}
\pa_{(\mu_1 \ldots}\pa_{\mu_{n-1}} g_{\mu_n)\tau} = \hlf
\pa_{\tau}\pa_{(\mu_1 \ldots}\pa_{\mu_{n-2}} g_{\mu_{n-1}\mu_n)} 
\end{equation}.}
The relation
\begin{equation}
\pa_{\mu}\Gam^{\tau}_{\nu\rho} = \ove{3} R_{\mu(\nu\rho)\tau}
\end{equation}
implies that
\begin{equation}
\pa_{\mu}\pa_{\rho}g_{\nu\tau} = - \ove{3} R_{\mu(\nu\tau)\rho}
\end{equation}
Each independent ordering in $S^{(4)}$ must combine to a double
derivative on $h_{ij}= g_{ij}-\eta_{ij}$ with both indices contracted
with the {\em dotted} $X^i$'s in order that we can find agreement with
the linearized result. This is the case if  
\begin{eqnarray}
a_1 = -\frac{\lam}{3} &~~~~~~~~~& a_2 = \frac{2\lam}{3} \non
b_1 = \frac{2\rho}{3} &~~~~~~~~~& b_2 = - \frac{\rho}{3} 
\end{eqnarray}
in which case $S^{(4)}$ reduces to 
\begin{equation}
S^{(4)} = \lam \pa_{i_3} \pa_{i_4}g_{i_1i_2} 
    \tr(\dot{{X}}^{i_1}\dot{X}^{i_2}{X}^{i_3}X^{i_4})
    + \rho \pa_{i_2}\pa_{i_4}g_{i_1i_3}
    \tr(\dot{{X}}^{i_1}X^{i_2}\dot{X}^{i_3}{X}^{i_4})
\label{lin4;2}
\end{equation}
Now if $\lam =2 \rho $ this combines to 
\begin{eqnarray}
S^{(4)} &=& \rho \pa_{i_3} \pa_{i_4}g_{i_1i_2} 
    \left(\tr(\dot{{X}}^{i_1}\dot{X}^{i_2}{X}^{i_3}X^{i_4})+
    \tr(\dot{{X}}^{i_1}X^{i_3}\dot{X}^{i_2}{X}^{i_4})
    +\tr(\dot{{X}}^{i_1}{X}^{i_3}{X}^{i_4}\dot{X}^{i_2}) \right)\non
&=& 3\rho \mbox{Str}(\dot{{X}}^{i_1}\dot{X}^{i_2}{X}^{i_3}X^{i_4}) 
\label{lin4;3}
\end{eqnarray}
Demanding consistency with the linearized result already yields a
unique solution 
\begin{equation}
3 \rho  = \frac{T_0}{2\, 2!}   ~~~\rar \rho = \frac{T_0}{2} \ove{6} 
\end{equation}
We still should check whether the Matrix-normal-coordinate conditions
are satisfied, i.e do the equations \rf{rnc} hold? Substituting the
solution  
\begin{eqnarray} 
a_1 = -\frac{2\rho}{3} &~~~~~~~~~& a_2 = \frac{4\rho}{3} \non
b_1 = \frac{2\rho}{3} &~~~~~~~~~& b_2 = - \frac{\rho}{3} 
\end{eqnarray}
we see that 
\begin{eqnarray}
2a_1+2b_1 = 0  &~~~~~&  a_1 + 4b_1 + 6b_2 = 0
\end{eqnarray}
vanish as required. 

Finally let us quickly check the consistency with the $U(1)$ result,
though this is implied by agreement with the linearized 
result. This requires that 
\begin{equation}
a_2 + b_1  = \frac{T_0}{2} \ove{3} 
\end{equation}
which is also seen to hold.

\subsection{The linearized coupling to all orders}

As preliminary,  note that 
the symmetrized trace, {\bf Str}, obeys the nice identity
\begin{equation}
\label{ld:eq7}  
\STr(ABC\ldots) = \Tr(A\,\Sym(BC\ldots))
\end{equation}
which may be proven using the expression 
\begin{equation}
  \STr(A^pB^q) = \frac{1}{(p+q)!}\Tr\frac{d^p}{d\alpha^p}\frac{d^q}{d\beta^q}(\alpha A+\beta B)^{p+q}|_{\alpha,\beta=0}
\end{equation}
Then the field equation which follows from the action
(\ref{ld:eq:1}) with  the substitution (\ref{ld:eq:2})
 equals 
\begin{eqnarray}
  \label{ld:eq:3}
  0&=&\Xdd_{\mu}-\dt\frac{\alp_n}{n!}\Sym(\Xd^jX^{k_1}\ldots
  X^{k_n})R_{k_1(\mu j)k_2\ldots k_n} \non
&& +\frac{n \alp_n}{2n!}\Sym(\Xd^i\Xd^jX^{k_1}\ldots
  X^{k_n})R_{k_1(ij)k_2\ldots k_{n-1}\mu} 
\end{eqnarray}
where the tensors are defined as\footnote{
Parenthetical symmetrization does not include a $1/n!$, viz.
$$
A_{(a}B_{b)} =A_aB_b+A_bB_a~~~~~~~~~~etc.
$$
Thus, symmetrized expressions have weight $n!$ instead of 1.}
\begin{equation}
  \label{ld:eq:4}
   R_{k_1(\mu j)k_2\ldots k_n}\equiv \frac{1}{n!}D_{(k_n}\ldots
   D_{k_3}R_{k1|\mu j|k_2)} + (\mu \leftrightarrow j)
\end{equation}
and are thus completely symmetric in the $k_i$ indices.
Notice that we keep $\alpha_n$ arbitrary. The value
of $\alpha_n$ should equal $(n-1)/(n+1)$ to have the right
$U(1)$ limit, but here we will present an alternative
derivation of this value based on diffeomorphism invariance.

\paragraph{{Field-Equation constraint}:}
\label{sec:field-equat-constr}
Checking that we really are in Riemann normal coordinates, we need to
show that the field equation has $X^i=tY^i$ at $t=1$ as solution. This is
obvious because one always ends up with a Riemann
tensor symmetrically contracted over three indices.

\paragraph{{Base-point transformation}:}
\label{sec:base-point-transf}
Next we need to find the basepoint transformation
\begin{equation}
  \label{ld:eq:5}
  X^i=\eps^i+t\Xd^i+\frac{t^2}{2}\Xdd^i +\ldots
\end{equation}
with $\Xd^i|_{t=0}=Z^i$. $\Xdd^i$ and higher derivatives are related to $Z^i$
through the field equation (\ref{ld:eq:3}).  We find
\begin{eqnarray}
  \label{ld:eq:6}
  \left(\dt\right)^{m+2}X_{\mu}&=&\frac{\alp_n}{n!}\left\{\left(\dt\right)^{m+1}\Sym(\Xd^jX^{k_1}\ldots
  X^{k_n})R_{k_1(\mu j)k_2\ldots k_n} \right. \non
&&\left.-\frac{n}{2}\left(\dt\right)^{m}\Sym(\Xd^i\Xd^jX^{k_1}\ldots
  X^{k_{n-1}})R_{k_1(i j)k_2\ldots k_{n-1}\mu}\right\} 
\end{eqnarray}
From this we see that if we
are only interested in terms linear in the Riemann tensor that we may ignore
multiple $d/dt$ derivatives on potential terms already involving
Riemann tensors. To remain linear in $\eps$ we need exactly one $X^i$ with the
remainder $\Xd^i$. Hence $n=m+2$. 
Using the identity for the symmetrized product (\ref{ld:eq7}) it is then
straightforward to show that
\begin{eqnarray}
  \label{ld:eq8}
  \left.\left(\dt\right)^{m+2}X_{\mu}\right|_{t=0}&=
  &\alp_{m+2}\left\{\eps^{k_1}\Sym(Z^jZ^{k_2}\ldots   
  Z^{k_{m+2}})R_{k_1(\mu j)k_2\ldots k_{m+2}} \right. \non
&&~~~\left.-\frac{\eps^{k_1}}{2}\Sym(Z^iZ^jZ^{k_2}\ldots
  Z^{k_{m+1}})R_{k_1(i j)k_2\ldots k_{m+1}\mu}\right\} 
\end{eqnarray}
Thus to linear order in both $\eps^i$ and Riemann tensors the coordinate
transformation from one set of Riemann normal coordinates
 to a nearby one is 
\begin{eqnarray}
  \label{ld:eq9}
  X_{\mu}&=&\eps_{\mu}+Z_{\mu}
  +\frac{\alp_n}{n!}\eps^{k_1}\left\{\Sym(Z^jZ^{k_2}\ldots 
  Z^{k_{n}})R_{k_1(\mu j)k_2\ldots k_{n}} \right. \non
&&\hspace{1in}\left.-\frac{1}{2}\Sym(Z^iZ^jZ^{k_2}\ldots
  Z^{k_{n-1}})R_{k_1(i j)k_2\ldots k_{n-1}\mu}\right\} 
\end{eqnarray}

To simplify this expression we notice that the symmetrization of the
$Z_i$ does not yet correspond with the properties of the 
tensor it is contracted
with. Let's rewrite
\begin{eqnarray}
  \label{ld:eq10}
  \Sym(Z^jZ^{k_2}\ldots
  Z^{k_{n}})R_{k_1(\mu j)k_2\ldots k_{n}}&=& \Sym(Z^{p_1}\ldots
  Z^{p_n})R_{k_1(\mu p_1)p_2\ldots p_n} \non
&=& \Sym(Z^{p_1}\ldots
  Z^{p_n})\left\{\frac{2}{n}D_{p_n}\ldots D_{p_3}R_{k_1(\mu p_1)p_2}
  \right. \non
&&~~~\left.+
  \frac{1}{n}\sum_{\ell=4}^{n+1}D_{p_n}\ldots
  {}_{p_{\ell}k_1p_{\ell-1}}\ldots D_{p_4}R_{p_2(\mu p_1)p_3}\right\} \non
  \non
&=& -\frac{1}{n}\Sym(Z^{p_1}\ldots
  Z^{p_n})R_{p_1(k_1 \mu)p_2 \ldots p_n}
\end{eqnarray}
Similarly
\begin{eqnarray}
  \label{ld:eq11}
  &&\Sym(Z^iZ^jZ^{k_2}\ldots Z^{k_{n-1}})R_{k_1ijk_2\ldots
  k_{n-1}\mu}=  \non
&&~~~~~~~~~~=\Sym(Z^{p_1}\ldots
  Z^{p_n})\left\{\frac{2(n-2)!}{n!}D_{p_n}\ldots
  D_{p_3}R_{k_1p_1p_2\mu} +\ldots \right\} \non
&&~~~~~~~~~~=\frac{1}{n(n-1)}\Sym(Z^{p_1}\ldots
  Z^{p_n})R_{p_1(k_1\mu)p_2\ldots p_n} 
\end{eqnarray}
This implies that the transformation $\Del X_{\mu} =
X_{\mu}(Z)-Z_{\mu}$ is simply
\begin{equation}
  \label{ld:eq12}
  \Del X_{\mu} =\eps_{\mu}-\frac{\alp_n}{(n-1)n!}\eps^{k_1}\Sym(Z^{p_1}\ldots
  Z^{p_n})R_{p_1(k_1\mu)p_2\ldots p_n}
\end{equation}

\paragraph{{Invariance of action}:}
\label{sec:incovariance-action}

Now we check whether the transformation of the action under the above
coordinate transformation can be cancelled by a covariant change of
the coupling constants: the Riemann tensors. The change in the action
equals $\Del X_{\mu}$ times the field equation in (\ref{ld:eq:3})
\begin{eqnarray}
  \label{ld:eq13}
  \Del S &=& \Tr\left(-\dt\Del
  X^{\mu}\left[\Zd_{\mu}-\frac{\alp_n}{n!}\Sym(\Zd^{\alp}Z^{k_1}\ldots
  Z^{k_n}) R_{k_1(\mu \alp)k_2\ldots k_n}\right]\right. \non
&& + \left.\Del
  X^{\mu}\left[\frac{n\alp_n}{2n!}\Sym(\Zd^{\alp}\Zd^{\bet}Z^{k_1}\ldots
  Z^{k_{n-1}})R_{k_1(\alp\bet)k_2\ldots k_{n-1}\mu}\right]\right)
\end{eqnarray}
It is straightforward to see that the single $d/dt$ derivative on $\Del
X_{\mu}$ yields
\begin{equation}
  \label{ld:eq14}
  \dt \Del X_{\mu} =
  -\frac{n\alp_n}{(n-1)n!}\eps^{k_1}\Sym(\Zd^{p_1}Z^{p_2}\ldots
  Z^{p_n})R_{p_1(k_1\mu)p_2\ldots p_n} 
\end{equation}
Then to linear order in $\eps^i$ and Riemann tensors the change in the action
is
\begin{eqnarray}
  \label{ld:eq15}
  \Del S &=&
  \frac{n\alp_n}{(n-1)n!}\eps^{k_1}\STr(\Zd^{\mu}\Zd^{p_1}Z^{p_2}\ldots
  Z^{p_n})R_{p_1(k_1\mu)p_2\ldots p_n} \non &&
  +\frac{n\alp_n}{2n!}\eps^{\mu}\STr(\Zd^{\alp}\Zd^{\bet}Z^{k_1}\ldots 
  Z^{k_{n-1}})R_{k_1(\alp\bet)k_2\ldots k_{n-1}\mu} \non
&=& \frac{n\alp_n}{n!}\eps^{k_1}\STr(\Zd^{\alp}\Zd^{\bet}Z^{p_2}\ldots
  Z^{p_n})\left(\frac{1}{(n-1)}R_{\bet (k_1\alp )p_2\ldots p_n}
  +\frac{1}{2}R_{p_2(\alp\bet )p_3\ldots p_n k_1}\right) 
\end{eqnarray}
At the end we expect that this reduces to
\begin{equation}
  \label{ld:eq16}
  \Del S \sim \eps^{k_1}D_{k_1} \Sym(\Zd^{\alp}\Zd^{\bet}Z^{p_1}
  \ldots)R_{p_1(\alp\bet)p_2\ldots} 
\end{equation}
Thus we have to somehow extract a single covariant derivative from the
symmetrized Riemann tensors.
Substituting their definition we find that
\begin{eqnarray}
  \label{ld:eq17}
  &&\STr(\Zd^{\alp}\Zd^{\bet}Z^{p_2}\ldots
  Z^{p_n})R_{\bet (k_1\alp )p_2\ldots p_n}= \non
&& \hspace{1in}=\STr(\Zd^{\alp}\Zd^{\bet}Z^{p_2}\ldots
  Z^{p_n})\left[\frac{2}{n}D_{p_n}\ldots D_{p_3}R_{\bet (k_1 \alp) p_2} +
  \right. \non
&&\hspace{2.5in}~~~~\left.\frac{1}{n}D_{p_n}\ldots {}_{\bet} \ldots
  D_{p_4}R_{p_2(k_1\alp)p_3}\right] 
\end{eqnarray}
and
\begin{eqnarray}
  \label{ld:eq18}
  &&\STr(\Zd^{\alp}\Zd^{\bet}Z^{p_2}\ldots
  Z^{p_n})R_{p_2(\alp\bet )p_3\ldots p_n k_1} =  \non
&&\hspace{1in}=
\STr(\Zd^{\alp}\Zd^{\bet}Z^{p_2}\ldots
  Z^{p_n})\left[\frac{2}{n}D_{p_n}\ldots D_{p_3}R_{k_1(\alp\bet)p_2} +
  \right. \non
&&\hspace{2.5in}~~~~\left.\frac{1}{n}D_{p_n}\ldots {}_{k_1}\ldots
  D_{p_4}R_{p_2(\alp\bet)p_3}\right]  
\end{eqnarray}
Combining these equations we find that 
\begin{eqnarray}
  \label{ld:eq19}
  \Del S &=&\frac{\alp_n}{n!}\eps^{k_1}\STr(\Zd^{\alp}\Zd^{\bet}Z^{p_2}\ldots
  Z^{p_n})\left(\frac{(n-2)}{(n-1)}D_{p_n}\ldots
  D_{p_3}R_{k_1(\alp\bet)p_2} +\right .\non 
&&  \hspace{2in}\frac{1}{2}D_{p_n}\ldots {}_{k_1}\ldots
  D_{p_4}R_{p_2(\alp\bet)p_3}+\non 
&& \left. \hspace{2in}\frac{2}{(n-1)}D_{p_n}\ldots {}_{\bet} \ldots
  D_{p_4}R_{k_1p_2p_3\alp}\right) 
\end{eqnarray}

Next we use the Bianchi identity to compare the three terms
\begin{equation}
  \label{ld:eq20}
  D_{p_3}R_{k_1\alp\bet p_2} = -D_{k_1}R_{\alp p_3\bet p_2}
  -D_{\alp}R_{p_3k_1\bet p_2}
\end{equation} 
One finds
\begin{eqnarray}
  \label{ld:eq21}
  \Del S &=& \frac{\alp_n}{n!}\eps^{k_1}\STr(\Zd^{\alp}\Zd^{\bet}Z^{p_2}\ldots
  Z^{p_n})\left(\frac{2(n-2)}{(n-1)}(-D_{p_n}\ldots D_{k_1}R_{\alp
  p_3\bet p_2}-D_{p_n}\ldots D_{\alp}R_{p_3k_1\bet p_2}) +\right .\non
&&   \hspace{2in}\frac{1}{2}D_{p_n}\ldots {}_{k_1}\ldots
  D_{p_4}R_{p_2(\alp\bet)p_3}+ \non
&& \left. \hspace{2in}\frac{2}{(n-1)}D_{p_n}\ldots {}_{\bet} \ldots D_{p_4}R_{k_1p_2p_3\alp}\right)
\end{eqnarray}

Finally, since we are only interested in  terms  linear in Riemann tensors
we may at this stage ignore all commutators of the covariant derivatives, and we end up
with solely 
\begin{eqnarray}
  \Del S &=& \frac{\alp_n}{n!}\eps^{k_1}\STr(\Zd^{\alp}\Zd^{\bet}Z^{p_2}\ldots
  Z^{p_n})\left(\frac{(n+1)(n-2)}{(n-1)}D_{k_1}D_{p_n}\ldots D_{p_4}R_{p_2\alp
  \bet p_3}\right) \non
&=& \frac{\alp_n(n-2)}{2n!}\frac{(n+1)}{(n-1)}\eps^{k_1}\STr(\Zd^{\alp}\Zd^{\bet}Z^{p_2}\ldots
  Z^{p_n})D_{k_1}R_{p_2(\alp
  \bet) p_3\ldots p_n}\label{ld:eq22}
\end{eqnarray}
which can be cancelled by suitable covariant derivatives of
the Riemann tensors
\begin{equation}
  \label{ld:eq23}
  \del S = -\frac{\alp_{n-1}}{2(n-1)!}\STr(\Zd^{\alp}\Zd^{\bet}Z^{p_2}\ldots
  Z^{p_n})\eps^{k_1}D_{k_1}R_{p_2(\alp
  \bet) p_3\ldots p_n}
\end{equation}
provided
\begin{eqnarray}
  \label{ld:eq24}
  \frac{\alp_{n-1}}{(n-1)!}&=&\frac{\alp_n}{n!}\frac{(n-2)(n+1)}{(n-1)}\non
\Leftrightarrow~~ \alp_{n-1}\frac{n}{n-2}&=&\alp_n\frac{n+1}{n-1}
\end{eqnarray}
with the obvious solution
\begin{equation}
  \label{ld:eq25}
  \alp_n=\frac{n-1}{n+1}
\end{equation}
as expected from the U(1) limit.

\section{The action to order $X^6$}

Here we present a two parameter family of actions that
satisfy the diffeomorphism constraint. The most general
action depends on 32 free parameters, we have chosen
the two parameters in such a way the action is 
relatively simple. Unfortunately, we have not succeeded in
finding values for the parameters for which the action can be
written in a nice compact form.

The action is obtained from (\ref{expan}) and is given by
\bea
g_{IJ} \dot{X}^{I} \dot{X}^{J} & = & 
\dot{X}^I \dot{X}^I  + \frac{1}{3}
R_{KIJL} X^I X^J \dot{X}^K \dot{X}^L \non
 & & + 
\frac{1}{6} \nabla_M
R_{KIJL} X^I X^J X^M \dot{X}^K \dot{X}^L \non
 & & + 
\frac{1}{20}\nabla_N \nabla_M
R_{KIJL} X^I X^J X^M X^N \dot{X}^K \dot{X}^L \non
 & & + 
\frac{2}{45} R_{KIJP} R_{PMNL}
X^I X^J X^M X^N \dot{X}^K \dot{X}^L 
\eea

The nonabelian generalization for the first three terms
is simply the maximally symmetrized version.
The last term becomes, using the curvature
tensor found in
section~4,
\be
\frac{2}{45} R_{kijp} R_{pmnl}
\tr {\rm Sym} (X^i X^j \dot{X}^k  )
{\rm Sym} ( X^m X^n  \dot{X}^l ) 
\ee
This term is uniquely determined and already not fully
symmetrically ordered.

The main problem is the one but last term. It 
has one contribution which is just
\be
\frac{1}{20}\nabla_n \nabla_m
R_{kijl} \tr {\rm Sym} 
(X^i X^j X^m X^n \dot{X}^k \dot{X}^l)
\ee
plus many $R^2$ terms. The latter depend on
32 free parameters. If we put 30 of those equal
to zero, in order to obtain a presentable expression,
we  find that the $R^2$ terms are of the form
\bea
\tr(X^n X^m X^i \dot{X}^j \dot{X}^k X^l ) & \times & 
T^1_{nmiklj} \nonumber \\
+ \tr(X^n X^m X^i \dot{X}^k X^l \dot{X}^j ) & \times & 
T^2_{nmiklj} \nonumber \\
+\tr(X^n X^i \dot{X}^k X^m X^l \dot{X}^j ) & \times &  
T^3_{nmiklj} 
\eea
with
\bea 
T^1_{nmiklj} & = & 
( -\frac{7}{120} - 2 \beta)     R_{m  k  l  p} R_{n  i  j  p}  
+ ( -\frac{7}{360} + \beta)     R_{m  l  k  p} R_{n  i  j  p}  \non
&&+ ( \frac{17}{180} + 2\beta)     R_{m  j  l  p} R_{n  i  k  p} 
+ ( -\frac{7}{90} - \beta)     R_{m  l  j  p} R_{n  i  k  p}  \non
&&+ (-\frac{1}{40} + \frac{\alp}{5} + \frac{\bet}{5})
     R_{j  k  l  p} R_{n  i  m  p} 
+ (\frac{1}{10} - \frac{3\alp}{5} + \frac{9\bet}{10})
     R_{j  l  k  p} R_{n  i  m  p} \non
\non &&+ (\frac{67}{360} + 4 \bet)
     R_{m  k  l  p} R_{n  j  i  p} 
+ (-\frac{7}{120} - 2 \bet)
     R_{m  l  k  p} R_{n  j  i  p} 
\non &&+ (\frac{1}{20} + \beta)
     R_{m  i  k  p} R_{n  j  l  p}
+ (\frac{1}{45} + \beta)
     R_{m  k  i  p} R_{n  j  l  p} 
\non &&+ (-\frac{11}{60} - 4\bet)
     R_{m  j  l  p} R_{n  k  i  p}
+ (-\frac{17}{180} + 2\bet)
     R_{m  l  j  p} R_{n  k  i  p} 
\non &&+ (\frac{13}{180} - \alp+\bet)
     R_{m  i  j  p} R_{n  k  l  p}
+ (-\frac{13}{360} + \frac{2\alp}{5} - \frac{3\bet}{5})
     R_{m  j  i  p} R_{n  k  l  p} 
\non &&+ (-\frac{4}{45} + \bet)
     R_{m  i  k  p} R_{n  l  j  p}
+ (\frac{1}{20} + \bet)
     R_{m  k  i  p} R_{n  l  j  p} 
\non &&+ (\frac{1}{120} + 4 \alp)
     R_{m  i  j  p} R_{n  l  k  p}
+ (\frac{13}{180} - \alp + \bet)
     R_{m  j  i  p} R_{n  l  k  p} 
\non &&+ (-\frac{1}{40} + \frac{\alp}{5} + \frac{\bet}{5} )
     R_{i  j  k  p} R_{n  l  m  p} 
+ (-\frac{3}{40} + \frac{2\alp}{5} - \frac{11 \bet}{10} )
     R_{i  k  j  p} R_{n  l  m  p} 
\non &&+ (-\frac{7}{360} - \frac{7 \alp}{5} - \frac{7 \bet}{5} )
     R_{j  k  l  p} R_{n  m  i  p} 
+ (-\frac{3}{40} + \frac{9\alp}{5} - \frac{6 \bet}{5} )
     R_{j  l  k  p} R_{n  m  i  p} 
\non &&+ (\frac{2}{45} + \frac{6\alp}{5} + \frac{6 \bet}{5} )
     R_{i  j  k  p} R_{n  m  l  p}
+ (-\frac{7}{360}  - \frac{3 \bet}{2} )
     R_{i  k  j  p} R_{n  m, l, p} \\
T^2_{nmiklj} & = & 
\frac{1}{180}
     R_{m  j  k  p} R_{n  i  l  p} 
 +\frac{7}{180}
     R_{m  k  j  p} R_{n  i  l  p}
\non && +(\frac{11}{360} - \frac{6\alp}{5} + \frac{3\bet}{10})
     R_{j  k  l  p} R_{n  i  m  p} 
 +(-\frac{11}{180} + \frac{9\alp}{5} - \frac{6\bet}{5})
     R_{j  l  k  p} R_{n  i  m  p}
\non && +\frac{5}{72}
     R_{m  i  l  p} R_{n  j  k  p}
 +\frac{1}{36}
     R_{m  l  i  p} R_{n  j  k  p} 
\non && +(-\frac{1}{9} + \alp-\bet)
     R_{m  i  k  p} R_{n  j  l  p}
 +(-\frac{19}{180} - \frac{2\alp}{5} - \frac{12\bet}{5})
     R_{m  k  i  p} R_{n  j  l  p} 
\non && +(-\frac{19}{180} - \frac{2\alp}{5} - \frac{12\bet}{5})
     R_{i  k  l  p} R_{n  j  m  p}
 +(\frac{29}{360} + \frac{2\alp}{5} + \frac{7\bet}{5})
     R_{i  l  k  p} R_{n  j  m  p} 
\non && +\frac{1}{36}
     R_{m  i  l  p} R_{n  k  j  p}
 +\frac{1}{36}
     R_{m  l  i  p} R_{n  k  j  p}  
\non && +(-\frac{67}{360} - 4\bet)
     R_{m  i  j  p} R_{n  k  l  p}
 +(\frac{23}{180} + 2\bet)
     R_{m  j  i  p} R_{n  k  l  p}  
\non && +(\frac{23}{180} + 2\bet)
     R_{i  j  l  p} R_{n  k  m  p}
 +(-\frac{7}{90} - \bet)
     R_{i  l  j  p} R_{n  k  m  p}  
\non && -\frac{2}{45}
     R_{m  j  k  p} R_{n  l  i  p}
 -\frac{2}{45}
     R_{m  k  j  p} R_{n  l  i  p}  
\non && +(\frac{1}{72} - \alp)
     R_{m  i  k  p} R_{n  l  j  p}
 +(\frac{29}{360} + \frac{2\alp}{5} + \frac{7\bet}{5})
     R_{m  k  i  p} R_{n  l  j  p}  
\non && +(\frac{7}{120} + 2 \bet)
     R_{m  i  j  p} R_{n  l  k  p}
 +(-\frac{7}{90} - \bet)
     R_{m  j  i  p} R_{n  l  k  p}  
\non && +\frac{1}{36}
     R_{i  j  k  p} R_{n  l  m  p}
 +\frac{1}{36}
     R_{i  k  j  p} R_{n  l  m  p}  
\non && +(-\frac{11}{180} + \frac{9\alp}{5} - \frac{6\bet}{5})
     R_{j  k  l  p} R_{n  m  i  p}
 +(\frac{11}{90} - \frac{18\alp}{5} + \frac{12\bet}{5})
     R_{j  l  k  p} R_{n  m  i  p}
\non && +(\frac{13}{60} - \frac{3\alp}{5} + \frac{17\bet}{5})
     R_{i  k  l  p} R_{n  m  j  p}
 +(-\frac{17}{180} + \frac{3\alp}{5} - \frac{7\bet}{5})
     R_{i  l  k  p} R_{n  m  j  p}
\non && +(\frac{7}{120} + 2 \bet)
     R_{i  j  l  p} R_{n  m  k  p}
 +(\frac{7}{36} - \bet)
     R_{i  l  j  p} R_{n  m  k  p}
 \non && -\frac{1}{18}
     R_{i  j  k  p} R_{n  m  l  p}
 -\frac{7}{72}
     R_{i  k  j  p} R_{n  m, l, p} \\
T^3_{nmiklj} & = & 
(-\frac{47}{120} - \frac{\alp}{5} - \frac{6\bet}{5})
     R_{m  i  k  p} R_{n  j  l  p}
 +(\frac{47}{360} + \frac{2\alp}{5} + \frac{12\bet}{5})
     R_{m  k  i  p} R_{n  j  l  p}
\non && -\frac{5}{144}
     R_{i  k  l  p} R_{n  j  m  p}
 -\frac{1}{72}
     R_{i  l  k  p} R_{n  j  m  p}
\non && +(\frac{17}{36} + \bet)
     R_{m  i  j  p} R_{n  k  l  p}
 +(-\frac{17}{180} - 2\bet)
     R_{m  j  i  p} R_{n  k  l  p}
\non && -\frac{5}{144}
     R_{i  j  l  p} R_{n  k  m  p}
 +\frac{7}{144}
     R_{i  l  j  p} R_{n  k  m  p}
\non && +(-\frac{1}{180} + \frac{\bet}{2})
     R_{m  i  k  p} R_{n  l  j  p}
 +(-\frac{47}{720} - \frac{\alp}{5} - \frac{6\bet}{5})
     R_{m  k  i  p} R_{n  l  j  p}
\non && +(-\frac{19}{360} - \frac{\bet}{2} )
     R_{m  i  j  p} R_{n  l  k  p}
 +(\frac{17}{360} + \bet)
     R_{m  j  i  p} R_{n  l  k  p}  
\non && -\frac{1}{72}
     R_{i  k  l  p} R_{n  m  j  p}
 -\frac{1}{72}
     R_{i  l  k  p} R_{n  m  j  p}  
\non && +\frac{7}{144}
     R_{i  j  l  p} R_{n  m  k  p}
 -\frac{11}{144}
     R_{i  l  j  p} R_{n  m  k  p}
\eea
It would be interesting to know if there
is a solution that can be written in an elegant compact
form that may suggest a generalization to higher orders.

\end{document}